%% file: ORB5Review.tex
\documentclass[preprint,10pt,fleqn]{elsarticle}%

\usepackage[dvipsnames]{xcolor}
\usepackage{amsmath}
\usepackage{amssymb}
\usepackage{pifont}
\usepackage{natbib}
\usepackage{fleqn}
\usepackage{amsfonts,bm}
\usepackage{verbatim}
\usepackage{graphicx}
\usepackage{epsfig}
\usepackage{float}
\usepackage{geometry}
\usepackage{lineno}
\usepackage{hyperref}
\usepackage{booktabs}
\usepackage{lipsum}
\usepackage{listings}
\usepackage{pgfplots}
\usepackage[utf8]{inputenc}
\usepackage{ifthen}

\newcommand{\figref}[1]{Fig.~\ref{#1}}

\newcommand{\dd}{{\rm d}}

\newcommand{\dt}{{\dd t}}
\newcommand{\ts}{\theta^{\star}}
\renewcommand{\vec}{\boldsymbol}
\newcommand{\grad}{{\vec{\nabla}}}
\newcommand{\dive}{{\grad}\cdot}
\newcommand{\rot}{{\grad}\times}

\newcommand{\pl}{\left(}
\newcommand{\pr}{\right)}
\newcommand{\bl}{\left[}
\newcommand{\br}{\right]}
\newcommand{\curl}{\left\{}
\newcommand{\curr}{\right\}}

\newcommand{\orb}{{\sc Orb5}}
\newcommand{\jac}{J}

\newcommand{\df}{{\rm{d}}}
\newcommand{\vc}[1]{{\bm{#1}}}
\newcommand{\gradperp}{\grad_\perp}
\newcommand{\pard}[2]{{\frac{\partial #1}{\partial #2}}}
\newcommand{\gav}[1]{\left\langle #1 \right\rangle}
\newcommand{\veps}{\varepsilon}
\newcommand{\Qpol}[1]{\mathcal{Q}_{s\text{#1}}^{\text{pol}}}
\newcommand{\Qgyr}[1]{\mathcal{Q}_{s\text{#1}}^{\text{gyr}}}
\newcommand{\Qdk}[1]{\mathcal{Q}_{\text{e#1}}^{\text{dk}}}
\newcommand{\nmin}{n_{\text{min}}}
\newcommand{\nmax}{n_{\text{max}}}
\newcommand{\mmin}{m_{\text{min}}}
\newcommand{\mmax}{m_{\text{max}}}

\newcommand{\bhat}{\widehat{\vec{b}}}
\renewcommand{\eqref}[1]{Eq.~(\ref{#1})}
\newcommand{\neqref}[1]{(\ref{#1})}
\newcommand{\ie}{i.e.\@}
\newcommand{\eg}{e.g.\@}

\DeclareMathOperator{\sign}{sign}
\DeclareMathOperator\erf{erf}
\journal{Computer Physics Communications}
\title{\orb{}: a global electromagnetic gyrokinetic code using the PIC approach in toroidal geometry}

\author[epfl]{E.~Lanti\corref{me}} \ead{emmanuel.lanti@epfl.ch}
\author[epfl]{N.~Ohana}
\author[garching]{N.~Tronko}
\author[garching]{T.~Hayward-Schneider}
\author[garching]{A.~Bottino}
\author[warwick]{B.~F.~McMillan}
\author[greifswald]{A.~Mishchenko}
\author[pppl]{A.~Scheinberg}
\author[garching]{A.~Biancalani}
\author[sib]{P.~Angelino}
\author[epfl]{S.~Brunner}
\author[pppl]{J.~Dominski}
\author[epfl]{P.~Donnel}
\author[epfl]{C.~Gheller}
\author[garching]{R.~Hatzky}
\author[cscs]{A.~Jocksch}
\author[epfl]{S.~Jolliet}
\author[garching]{Z.X.~Lu}
\author[garching]{J.~P.~Martin~Collar}
\author[garching]{I.~Novikau}
\author[garching]{E.~Sonnendrücker}
\author[epfl]{T.~Vernay}
\author[epfl]{L.~Villard}

\cortext[me]{Corresponding author: E. Lanti; Email: emmanuel.lanti@epfl.ch}
\address[epfl]{\'Ecole Polytechnique F\'ed\'erale de Lausanne (EPFL), Swiss Plasma Center
  (SPC), CH-1015 Lausanne, Switzerland}
\address[garching]{Max-Planck-Institut für Plasmaphysik, D-85748 Garching, Germany}
\address[greifswald]{Max-Planck-Institut für Plasmaphysik, D-17491 Greifswald, Germany}
\address[warwick]{CFSA, Department of Physics, University of Warwick, Coventry CV4 7AL, United Kingdom}
\address[pppl]{Princeton Plasma Physics Laboratory, Princeton, New Jersey 08540, USA}
\address[sib]{Swiss Institute of Bioinformatics (SIB), Lausanne, Switzerland}
\address[cscs]{CSCS, Swiss National Supercomputing Centre, Via Trevano 131, 6900 Lugano,
  Switzerland}
\begin{document}
\begin{frontmatter}
  \begin{abstract}
    This paper presents the current state of the global gyrokinetic code \orb{} as an update of the
    previous reference [Jolliet \textit{et al.}, Comp. Phys. Commun. \textbf{177} 409 (2007)]. The
    \orb{} code solves the electromagnetic Vlasov-Maxwell system of equations using a PIC scheme and
    also includes collisions and strong flows. The code assumes multiple gyrokinetic ion species at
    all wavelengths for the polarization density and drift-kinetic electrons. Variants of the
    physical model can be selected for electrons such as assuming an adiabatic response or a
    ``hybrid'' model in which passing electrons are assumed adiabatic and trapped electrons are
    drift-kinetic. A Fourier filter as well as various control variates and noise reduction
    techniques enable simulations with good signal-to-noise ratios at a limited numerical cost. They
    are completed with different momentum and zonal flow-conserving heat sources allowing for
    temperature-gradient and flux-driven simulations. The code, which runs on both CPUs and GPUs, is
    well benchmarked against other similar codes and analytical predictions, and shows good
    scalability up to thousands of nodes.
  \end{abstract}
\end{frontmatter}

%\ifdefined\review \linenumbers \fi \ifdefined\preprint \linenumbers \fi
\tableofcontents
\newpage

%\printnomenclature

\input{sections/introduction}
\input{sections/gk_model}
\input{sections/numerical_implementation}
\input{sections/results}
\input{sections/conclusion}
% -------------------------------------------------------------------------------
% BIBLIOGRAPHY
% -------------------------------------------------------------------------------
\section*{References}

\bibliographystyle{elsarticle-num} % or unsrt
\bibliography{ORB5Review.bib}
\end{document}

%% file: sections/introduction.tex
\section{Introduction}
\label{sec:introduction}

Understanding the critical phenomena limiting the performance of magnetic confinement devices is
crucial to achieve a commercially viable fusion energy production. Among them, microinstabilities
play a key role as they are closely linked to the tokamak confinement properties. For example,
turbulent transport induced by microinstabilities mainly governs the heat and particle losses in
toroidally confined plasmas. Another important issue is the interaction between waves and energetic
particles produced by the fusion process or resulting from the application of heating by neutral
beam injection (NBI) or ion cyclotron range of frequencies (ICRF). In this case, the energetic
particles interact with the bulk plasma and destabilize various eigenmodes of the shear Alfv\'en
wave such as toroidal Alfv\'en eigenmodes (TAE) or the energetic particle modes (EPM), which
deteriorate the confinement properties.

It is shown both experimentally \cite{Fonck1993,Zweben2007} and theoretically \cite{Horton1999} that
these drift-wave-type microinstabilities as well as Alfv\'en eigenmodes \cite{Cheng1985,Chen2016}
have a low frequency compared to the ion gyro frequency of strongly magnetized plasmas and are of
small amplitude in the core region compared to the background quantities. This motivates the use of
gyrokinetic theory \cite{Brizard2007,Tronko2017a} which retains a kinetic description of the problem
while reducing the numerical cost for solving the equations by removing the fast gyro angle
dependence of the system in a consistent way and thus reducing the phase space dimensionality from
6D to 5D.

Among the three main numerical approaches used to solve the gyrokinetic equations \cite{Garbet2010}:
Lagrangian \cite{Jolliet2007,Idomura2003,Lin1998a,Ku2018,Wang2007,Dimits1996,Korpilo2016}, Eulerian
\cite{Jenko2000a,Idomura2008,Candy2003a,Peeters2004,Watanabe2006a,Kotschenreuther1995}, and
Semi-Lagrangian \cite{Grandgirard2016}, the Lagrangian particle-in-cell (PIC) scheme
\cite{Birdsall2004} was the first introduced in the context of gyrokinetic simulations
\cite{Lee1987}. It consists of initially sampling the phase space using numerical particles, also
called markers, that represent a portion of the phase space and following their orbit in the 5D
space.

The \orb{} code is a nonlinear global PIC code used for solving the gyrokinetic Vlasov-Maxwell
system accounting for the presence of collisions and sources. It is based on a 3D finite element
representation of the fields using B-spline basis functions up to third order. It uses toroidal
magnetic coordinates and a field-aligned Fourier filter which drastically reduces particle
noise. Originally presented in \cite{Tran1999} and further developed in \cite{Jolliet2007} for the
electrostatic (ES) and collisionless limit, the \orb{} code has since undergone a substantial amount
of additions. Those improvements are targeting the physical models, with \eg{} drift-kinetic
electron dynamics, electromagnetic (EM) perturbations \cite{Bottino2011}, multiple gyrokinetic ion
species, inter and intraspecies collisions \cite{Vernay2010}, hybrid electron model
\cite{Jolliet2009a,Vernay2013a}, removal of the long wavelength approximation \cite{Dominski2017},
various heating sources \cite{McMillan2008,Jolliet2009b} and strong flows \cite{Collier2016}, and
the numerical side with \eg{} the enhanced control variate
\cite{Mishchenko2017,Hatzky2007,Hatzky2019}, and, more recently, the mixed-representation
``pullback'' scheme \cite{Mishchenko2018a} resolving the so-called cancellation problem for EM
simulations, various noise control operators (generalized moment-conserving Krook operator
\cite{McMillan2008}, coarse graining \cite{Vernay2012}, and quadtree \cite{Sonnendrucker2015}), and
a thorough refactoring with multithreading using OpenMP and OpenACC which will be detailed in a
separate publication. The aim of this paper is to review these improvements, and present the current
status of the code and illustrate its performance and capabilities with a few significant results.

The present paper is organized as follows. Section \ref{sec:model} presents the gyrokinetic model
implemented in \orb{}. Section \ref{sec:numericalimplementation} describes the numerical
implementation of the gyrokinetic equations as well as the numerical methods used in the code. The
parallel efficiency and a few illustrative, physically relevant simulation results are presented in
Section \ref{sec:results}. Finally, Section \ref{sec:conclusion} presents the conclusions and future
work.

%%% Local Variables:
%%% mode: latex
%%% TeX-master: "../ORB5Review"
%%% End:

%% file: sections/gk_model.tex
\section{Gyrokinetic model}
\label{sec:model}
\subsection{Magnetic geometry, coordinate system, and normalization}
\label{sec:definitions}
The background fields of a tokamak are usually approximated as axisymmetric. A general axisymmetric
magnetic field in the nested-flux-surface region may be
expressed as
\begin{equation}
  \vec{B}(\psi) = F(\psi)\grad\varphi + \grad\psi \wedge \grad\varphi,
\end{equation}
where $F(\psi)$ is the poloidal current flux function, $\psi$ is the poloidal magnetic flux and
$\varphi$ is the toroidal angle. The \orb{} code uses ideal-MHD equilibria, solution of the
Grad-Shafranov equation, that are produced by the CHEASE code \cite{Lutjens1996}. It can also use an
analytical \textit{ad-hoc} magnetic equilibrium comprising circular concentric magnetic surfaces.

A straight-field-line coordinate system is used in \orb{}. The magnetic surfaces are labeled by
$s = \sqrt{\psi/\psi_{\text{edge}}}$ where $\psi_{\text{edge}}$ is the value of $\psi$ at the radial
edge, the toroidal angle is $\varphi$, and the poloidal angle is defined by
\begin{equation}
  \ts = \frac{1}{q(s)}\int^\theta_0
  \frac{\vec{B}\cdot\grad\varphi}{\vec{B}\cdot\grad\theta'}\dd\theta',
\end{equation}
where $q(s)$ is the safety factor profile and $\theta$ is the geometric poloidal angle.

All the physical quantities in \orb{} are normalized according to four reference parameters; these
normalizations are used internally and in the code output. The reference quantities are the ion mass
$m_\text{i}$, the ion charge $q_\text{i}=eZ_\text{i}$ with $e$ being the elementary charge and
$Z_\text{i}$ the ion atomic number, the magnetic field amplitude at the magnetic axis $B_0$, and the
electron temperature $T_\text{e}(s_0)$ at a reference magnetic surface $s_0$. Note that for
simulations with multiple ion species, the user must define a reference ion species for the
normalization. Derived units are then defined with respect to these four parameters: time is in
units of the inverse of the ion cyclotron frequency
$\Omega_{\text{ci}} = q_\text{i}B_0/m_\text{i} c$ with $c$ the speed of light in vacuum (CGS units
are used in this paper), velocities are normalized to the ion sound velocity
$c_s = \sqrt{eT_\text{e}(s_0)/m_\text{i}}$, lengths are given in units of the ion sound Larmor
radius $\rho_s = c_s/\Omega_{\text{ci}}$, and the densities are normalized to the volume averaged
density $\bar{n}$. These reference quantities are then used to construct normalizations for other
quantities in the code such as the electrostatic potential, various fluxes, etc.

\subsection{GK Equations for fields and particles}
\label{sec:gkeq}
%{\color{blue} The conversion to SI units is done accordingly to the rules: $e=q_s$, $m=m_s$, $c=1$,
  %$8\pi=2\mu_0$, $\mu=m_s \mu$}

The gyrokinetic Vlasov-Maxwell model implemented in \orb{} is derived from variational principles
\cite{Sugama2000, Tronko2016} which have some advantages with respect to the models implemented in
other gyrokinetic codes which are derived outside of a structural framework. The first advantage is
the possibility to include all necessary approximations into the expression of the action before
deriving the equations of motion. The second advantage consists of the possibility to consistently
derive exactly conserved quantities, corresponding to the model, such as the energy. In the \orb{}
code these quantities are then used for diagnostics and the verification of the quality of the
simulations. Finally, the variational formulation directly provides the weak form of gyrokinetic
Poisson and Ampère equations suitable for a finite element discretization.

The choice of the ordering plays a crucial role in defining the complexity of the gyrokinetic model,
and in particular the nonlinear terms which are taken into account. The gyrokinetic variational
principle corresponding to the \orb{} model is established according to the specific gyrokinetic
ordering suitable for numerical implementation. In particular, it means that all the geometrical
effects due to the non-uniformity of the background magnetic field are considered one order smaller
than the relative fluctuations of the electromagnetic fields. To quantify that statement, we define
the magnetic field geometry-related small parameter $\epsilon_B=\rho_{\mathrm{th}}/L_B$, where
$\rho_{\mathrm{th}}$ is the thermal Larmor radius of the particle and $L_B=|\bm\nabla B/B|^{-1}$
sets up the length scale of the background magnetic field variation. The
electromagnetic-fluctuations-related small parameter is defined by
$\epsilon_{\delta} \sim|\vec{B}_1|/B\sim c|\vec{E}_{1\perp}|/(B
v_{\mathrm{th}})\sim\left(k_{\perp}\rho_{\mathrm{th}}\right)\
e\phi_1/T_\text{i}\equiv\epsilon_{\perp}e\phi_1/T_\text{i}$, where $\vec{E}$ is the electric field,
$v_{\mathrm{th}}$ is the thermal velocity, $k_\perp$ is the wave number perpendicular to the
magnetic field $B$ the amplitude of the background magnetic field, $\phi_1$ is the perturbed
electrostatic potential, $T_\text{i}$ is the ion temperature, the subscript 1 refers to the
fluctuating part of the corresponding fields and the subscript $\perp$ represents the component
perpendicular to the magnetic field line. The parameter $\epsilon_{\perp}$ allows the distinction
between the gyrokinetic theory with $\epsilon_{\perp}\sim \mathcal{O}(1)$ and the drift-kinetic
theory with $\epsilon_{\perp}\ll 1$. Both type of models are implemented in the code \orb{}.

As shown in \cite{Tronko2016, Tronko2017} the ordering $\epsilon_B=\epsilon_{\delta}^2$ corresponds
to gyrokinetic models implemented in most global codes. In particular, it has been demonstrated that
the \orb{} equations can be derived via variational calculation from the second order with respect
to the parameter $\epsilon_{\delta}$ field-particle Lagrangian. Below we present the variational
framework and summarize the main gyrokinetic equations resulting from the variational derivation.

The expression of the action functional leading to the \orb{} code gyrokinetic Maxwell-Vlasov
equations containing first order geometric corrections, \ie{} $\mathcal{O}(\epsilon_B)$
terms, and the electromagnetic corrections up to the second order, \ie{}
$\mathcal{O}(\epsilon_{\delta}^2)$ terms, is given by:

\begin{eqnarray}
  \label{L_ORB5}
  \mathcal{A}=\int_{t_0}^{t_1}\ \dt\ \mathcal{L}
  &=&
      \sum_{s} \int \dt\ \dd \Omega
      \left(\frac{q_s}{c} \vec{A}^*\cdot\dot{\vec{X}}+\frac{m_s c}{q_s}\mu\dot{\theta}-H_0\right)\ f_s
  \\
  \nonumber
  &-&
      \epsilon_{\delta}\sum_{s\neq\text{e}} \int \dt\ \dd \Omega\ H_1\ f_s
      -\epsilon_{\delta} \int \dt\ \dd \Omega\ H_1^{\mathrm{dk}}f_\text{e}
  \\
  \nonumber
  &-&\epsilon_{\delta}^2\sum_{s\neq \text{e}} \int\ \dt\  \dd \Omega \ H_{2} f_{\text{eq,s}}
      -\alpha\epsilon_{\delta}^2\int \dt\ \dd \Omega\ H_2^{\mathrm{dk}}\ f_{\text{eq,e}}
      -\alpha\epsilon_{\delta}^2\int\ \dt\ \dd V \ \frac{\left|\bm\nabla_{\perp} A_{1\|}\right|^2}{8\pi},
      \nonumber
\end{eqnarray}
where $\alpha=0$ corresponds to the electrostatic model and $\alpha=1$ to the electromagnetic model,
$\dd \Omega = \dd V \dd W$ with $\dd V=d^3{\vec X}$ and $\dd W=B_{\|}^*\dd\mu\ \dd p_{z}$
represents the infinitesimal volume of the reduced (gyrocenter) phase space, $B_{\|}^*$ is defined
as the parallel component of the symplectic magnetic field
$\vec{B}^*=\bm\nabla\times{\vec A}^*$ with ${\vec A}^*={\vec A}+(c/q_s)\ p_{z} \bhat$
being the symplectic magnetic potential and $\bhat$ being the unit vector parallel to the magnetic
field line. The action is derived using the $p_{z}$ formulation in which we define the reduced
gyrocenter position $\vec{X}$, the canonical gyrocenter momentum
$p_{z}=m_s v_{\|}+\alpha\ \epsilon_{\delta} (q_s/c) A_{1\|}$ with the parallel velocity $v_\|$, the
magnetic moment $\mu$ and the fast gyro angle $\theta$. The sums are made over all the species $s$
except for the second and third sums where the electrons are excluded because they are treated as
drift-kinetic. The first and the second terms of the gyrokinetic action are gyrocenter contributions
and the last term is a contribution from the perturbed magnetic field.

Before presenting the equations of motion implemented in \orb{}, we discuss all necessary
approximations included in the gyrokinetic action given by \eqref{L_ORB5}.  The first three terms of
the action involves the full distribution functions $f_s$, while the fourth and fifth terms,
involving the nonlinear Hamiltonian $H_{2}$, involve equilibrium distribution functions
$f_{\text{eq},s}$, which are by definition invariant under the unperturbed Hamiltonian dynamics,
\ie{} they satisfy the condition $\{f_{\text{eq},s},H_0\}=0$. This approximation brings several
simplifications in the model. First, it results in the linearization of the gyrokinetic Poisson and
Amp\`ere equations. Second, it simplifies the gyrokinetic Vlasov equation by excluding some
nonlinear terms from the gyrocenter characteristics associated with the Hamiltonian $H_2$.

The gyrocenter model is fixed via the Hamiltonians $H_0$ (non-perturbed dynamics), $H_1$ (linear
gyrocenter dynamics), $H_1^{\mathrm{dk}}$ (linear drift-kinetic dynamics for electrons), and $H_2$
(nonlinear second order gyrocenter dynamics). The choice of the linear $H_1$, $H_1^{\mathrm{dk}}$
and nonlinear Hamiltonians $H_2$ determines the expressions for the gyrokinetic charge and current
in the reduced Poisson and Ampère equations. In this section we present the general electromagnetic
model of the \orb{} code. For further options and approximations implemented on the level of the
reduced particle dynamics, see the sections below.

Concerning the field part of gyrokinetic action, three approximations have been made. First of all,
the quasi-neutrality approximation, which allows one to neglect the perturbed electric field energy
$-\epsilon_{\delta}^2\int\ \dt\ \dd V \ \left|\vec{E}_{1}\right|^2/8\pi$. The second
approximation consists in neglecting the magnetic compressibility of perturbations with
$B_{1\|}=\epsilon_{\delta} |\vec B_{1\perp}|$, \ie{} the parallel component of the perturbed
magnetic field is neglected and only the perpendicular part of the perturbed magnetic field
$\vec{B}_{1\perp}=\bhat\times\bm\nabla A_{1\|}$, associated with $A_{1\|}$, is
implemented. Finally, due to the chosen ordering, the background component of the magnetic field can
be excluded from the Maxwell part of the gyrokinetic action.

The background Hamiltonian contains information about the kinetic energy of a charged particle
moving in a magnetic field with amplitude $B$:
\begin{equation}
H_0= \frac{p_{z}^2}{2 m_s}+\mu B.
\label{H_0}
\end{equation}
The linearized Hamiltonian model for ions is given by the gyroaveraged linear electromagnetic potential:
\begin{equation}
H_1=q_s\left\langle\phi_1-\alpha A_{1\|}\frac{p_{z}}{m_sc} \right\rangle,
\label{H_1}
\end{equation}
where $\left\langle\dots\right\rangle$ is the gyroaveraging operator.  The gyroaveraging is removed
from the linear Hamiltonian model for the electrons which are considered as drift-kinetic:
\begin{equation}
H_1^{\mathrm{dk}}=-e\left(\phi_1(\vec{X})-\alpha A_{1\|}(\vec{X})\frac{p_{\text{z}}}{m_\text{e}c} \right).
\label{H_1_FLR}
\end{equation}
The nonlinear Hamiltonian model which contains all orders in finite Larmor radius (FLR) in its
electrostatic part and up to second order FLR terms in its electromagnetic part is considered for
ions only:
\begin{eqnarray}
\label{H_2}
H_{2}&=&-\frac{q_s^2}{2B}\frac{\partial}{\partial\mu}\left\langle\widetilde{\phi}_1\left({\vec{X}+
\bm\rho_0}\right)^2\right\rangle
\\
&+&\alpha \frac{q_s^2}{2 m_s c^2}\bl A_{1{\|}}(\vec{X})^2+m_s\left(\frac{c}{q_s}\right)^2 \frac{\mu}{B}A_{1\|}(\vec{X})\bm\nabla_{\perp}^2 A_{1\|}\left(\vec{X}\right)\br,
\nonumber
\end{eqnarray}
where $\widetilde{\phi}_1$ represents the fluctuating part of a perturbed electrostatic potential
and $\vec{\rho}_0$ is the lowest order guiding-center displacement. Finally the second order
Hamiltonian for the electrons contains the first FLR correction to the electromagnetic potential
only:
\begin{eqnarray}
\label{H_2_dk}
H_{2}^{\mathrm{dk}}=\alpha \frac{e^2}{2 m_\text{e} c^2}\ A_{1{\|}}(\vec{X})^2.
\nonumber
\end{eqnarray}

\subsubsection{Quasineutrality and Ampère equations}
\label{sec:qsea}
The corresponding quasineutrality equation in a weak form is derived from the gyrokinetic action,
\eqref{L_ORB5}:
\begin{align}
  \label{quasin_full}
  &\sum_{s\neq\text{e}}\Qgyr{}+\Qdk{} = \sum_{s\neq\text{e}} \Qpol{},\\
  \label{gyr_full}
  &\Qgyr{} = \int \dd \Omega\ f_s\ q_s\left\langle\widehat{\phi}_1\right\rangle,\\
  \label{dk_full}
  &\Qdk{} = -\int \dd \Omega\ f_\text{e}\ e\,\widehat{\phi}_1(\vec{X}),\\
  \label{pol_full}
  &\Qpol{} = \epsilon_{\delta}\int \dd \Omega\ f_{\text{eq},s}\ \frac{q_s^2}{B}\frac{\partial}{\partial\mu}\left(\left\langle\phi_1\widehat{\phi}_1\right\rangle-\Big\langle\phi_1\Big\rangle\left\langle\widehat{\phi}_1\right\rangle\right),
\end{align}
where $\widehat{\phi}_1$ represents an arbitrary test function, which can be a B-spline of a
required order for the finite element discretization. On the left-hand side of the equation,
$\Qgyr{}$ is associated with the gyro-charge of the ions, $\Qdk{}$ with the drift-kinetic charge of
the electrons and on the right-hand side, $\Qpol{}$ is associated with the linear ion polarization
charge. Note that due to the drift-kinetic approximation used for the electrons, there is no linear
contribution to the polarization density from the electron species.

Similarly, the Ampère equation issued from the variational principle is given by
\begin{eqnarray}
\label{GK_Ampere_full}
0=&-&\epsilon_{\delta}\int \frac{\dd V}{4\pi}\
\bm\nabla_{\perp} A_{1\|}\cdot\bm\nabla_{\perp}\widehat{A}_{1\|}
+
\sum_{s\neq\text{e}}\int \ \dd \Omega\ f_s \
\frac{q_sp_{z}}{m_s c}\left\langle\widehat{A}_{1\|}\right\rangle-
\int \ \dd \Omega\ f_\text{e} \
\frac{e\,p_{z}}{m_\text{e} c}\ \widehat{A}_{1\|}
\\
\nonumber
&-&\epsilon_{\delta}\int \dd \Omega\ f_{\text{eq,e}}\ \left(\frac{e^2}{m_\text{e} c^2}A_{1\|}\widehat{A}_{1\|}
\right)
\nonumber
\\
&-&\sum_{s\neq \text{e}}\epsilon_{\delta}\int \dd \Omega\ f_{\text{eq},s}\ \bl\frac{q_s^2}{m_s c^2}A_{1\|}\widehat{A}_{1\|}
+\frac{\mu}{2B}
\pl
A_{1\|}\bm\nabla_{\perp}^2\widehat{A}_{1\|}+
\widehat{A}_{1\|}\bm\nabla_{\perp}^2A_{1\|}
\pr
\br,
\nonumber
\end{eqnarray}
for all test functions $\widehat{A}_{1\|}$.

\subsubsection{Nonlinear gyrokinetic Vlasov equation}
The gyrokinetic Vlasov equation for the distribution function $f_s$ of each species $s$ is
reconstructed from the linearized gyrocenter characteristics according to the approximations
performed on the action functional given by \eqref{L_ORB5}:
\begin{eqnarray}
0=\frac{\dd f_s}{\dd t}=\frac{\partial f_s}{\partial t}
+\dot{\vec X}\cdot\bm\nabla f_s+\dot{p_{z}}\frac{\partial f_s}{\partial p_{z}},
\label{eq:Vlasov_general}
\end{eqnarray}
where the gyrocenter characteristics depend on the linearized Hamiltonian model:
\begin{align}
  \label{eq:xdot}
  \dot{\vec X}&=\frac{c\bhat}
                     {q_s B_{\|}^{*}}\times\bm\nabla H+\frac{\partial
                     H}{\partial{{p}_{z}}}\frac{\vec{B}^{*}}{B_{\|}^{*}},\\
  \label{eq:pdot}
  \dot{p}_{z}&=-\frac{\vec{B}^{*}}{B_{\|}^{*}}\cdot \bm\nabla H,
\end{align}
with $H=H_{0} +\epsilon_{\delta}H_1$, where $H_0$ is a Hamiltonian corresponding to the
non perturbed guiding-center dynamics given by \eqref{H_0} and $H_1$ corresponds to the first
order gyrocenter contributions given by \eqref{H_1}.

For the ordering considered above, the characteristics become:
\begin{align}
  \label{linear_charakt}
  \dot{\vec X}&=\frac{c\bhat}
                   {q_s B_{\|}^{*}}\times\bm\nabla\bl\mu B+\epsilon_{\delta} q_s\left(
                   \left\langle\phi_1\right\rangle-\alpha\frac{p_{z}}{m_s} \left\langle
                   A_{1\|}\right\rangle\right)\br+\frac{\vec{B}^{*}}{B_{\|}^{*}}\left(\frac{p_{z}}{m_s}-\epsilon_{\delta}\alpha\frac{q_s}{m_s}
                   \left\langle A_{1\|}\right\rangle\right),\\
  \label{eq:pdot_exp}
\dot{p}_{\text{z}}&=-\frac{\vec{B}^{*}}{B_{\|}^{*}}\cdot \bm\nabla \bl \mu B+\epsilon_{\delta} q_s\left(
             \left\langle\phi_1\right\rangle-\alpha\frac{p_{z}}{m_s}\left\langle A_{1\|}\right\rangle\right)\br,
\end{align}
which can be written in a different form to make the usual drift velocities appear:
\begin{align}
  \dot{\vec X}=& \frac{p_{z}}{m_s}\bhat -
                    \frac{c\,p_{z}^2}{q_sm_s}\frac{1}{B^*_\|}\bl\bhat\times \pl \widehat{\vec
                    b} \times \frac{\bm\nabla\times{\vec B}}{B}\pr\br + \frac{c}{q_sB^*_\|}\pl
                    \mu B+\frac{p_{z}^2}{m_s}\pr\widehat{\vec
                    b}\times\frac{\bm\nabla B}{B} \\
  \nonumber&+\epsilon_\delta \frac{c}{B^*_\|}\bhat\times\grad\pl
             \left\langle\phi_1\right\rangle - \alpha\frac{p_{z}}{m_s}\left\langle A_{1\|}\right\rangle\pr
             +\epsilon_\delta\alpha \frac{c\,p_{z}}{m_sB^*_\|}\left\langle A_{1\|}\right\rangle{\boldsymbol
             \kappa}-\epsilon_\delta\alpha \frac{q_s}{m_s}\left\langle A_{1\|}\right\rangle\bhat\\
  \equiv& {\vec v_\|} + {\vec v_\text{D}}+{\vec v_{\grad B}}+{\vec
          v_{\text{C}}}+{\vec v_{E\times B}}+{\vec v_{A_\|}},
\end{align}
where $\boldsymbol \kappa$ is the curvature vector
\begin{equation}
  \vec{\kappa}=\bhat\times\left[\bhat\times\frac{\vec{\nabla}\times\vec{B}}{B}\right]+\frac{\vec{\nabla}{\vec{B}}\times\bhat}{B}.
\end{equation}

The first term of the equation is the parallel velocity ${\vec v_\|}$, the second is the
diamagnetic drift ${\vec v_\text{D}}$, the third term can be separated in the $\bm\nabla B$ drift
${\vec v}_{\grad B}$ and curvature drift ${\vec v_\text{C}}$, the fourth is the $E\times B$
drift ${\vec v_{E\times B}}$, and the last two terms are labeled as ${\vec v_{A_\|}}$. Similarly,
the same procedure can be applied to the $p_{z}$ characteristic:
\begin{align}
  \dot{p_{z}}=&\mu B\,\dive{\vec{B}} - \frac{c\,p_{z}\mu}{q_s B^*_\|}\bl \bhat \times
              \pl\bhat\times\frac{\rot{\vec{B}}}{B}\pr\br \cdot \grad B \\
  & - \epsilon_\delta \grad \pl\left\langle\phi_1\right\rangle - \alpha\frac{p_{z}}{m_s}\left\langle
    A_{1\|}\right\rangle \pr \cdot \pl q_s \bhat +
    \frac{c\,p_{z}}{B^*_\|}\vec{\kappa} \pr \\
  \equiv & -\frac{m_s}{p_{z}}\pl{\vec v_\|} + {\vec v_\text{D}}+{\vec v_{\text{C}}}\pr\cdot\grad\pl\mu B +
           \epsilon_\delta q_s\left\langle\phi_1\right\rangle - \epsilon_\delta \alpha\frac{p_{z}}{m_s}\left\langle A_{1\|}\right\rangle\pr.
\end{align}

In the \orb{} gyrokinetic model, different additional approximations can be made on the total time
derivative operator introduced in \eqref{eq:Vlasov_general}: the linear and/or neoclassical
limits. To this end, the characteristic equations \neqref{linear_charakt} and \neqref{eq:pdot_exp}
are slightly modified. In the linear limit, all the perturbed terms, proportional to
$\epsilon_\delta$, are neglected leading to:
\begin{align}
  \dot{\vec X}^{\text{lin}}&={\vec v_\|} + {\vec v_\text{D}}+{\vec v_{\grad B}}+{\vec v_\text{C}},\\
  \dot{p}_{z}^{\text{lin}}&=-\mu\frac{m_s}{p_{z}}\pl{\vec v_\|} + {\vec v_\text{D}}+{\vec
                            v_{\text{C}}}\pr\cdot\grad B.
\end{align}

The neoclassical limit is made neglecting the electromagnetic fields and assuming small banana
widths as compared to the characteristic lengths of the system which leads to neglecting all drift
velocities compared to the parallel drift velocity:
\begin{align}
  \dot{\vec X}^{\text{neo}}&={\vec v_\|} ,\\
\dot{p}_{z}^{\text{neo}}&=-\mu\frac{m_s}{p_{z}}{\vec v_\|}\cdot\grad B.
\end{align}

\subsection{Variants of the physical models}
In this section, we present the different variants of the physical model presented above that are
available in the \orb{} code. Usually, each variant can be obtained in the framework of the
variational formulation by changing the $H_0$, $H_1$ and $H_2$ Hamiltonians according to the
corresponding approximations. This is the case for the long-wavelength approximated electromagnetic
model as well as the electrostatic models with a Padé approximation and a strong background
flow. For the adiabatic electron model, an external coupling of the gyrokinetic equations with a
fluid polarization density of the electrons is assumed. Including this model into the general
framework requires some additional approximations on the field term of the field-particles
Lagrangian given by \eqref{L_ORB5}. Note that these models are not necessarily mutually exclusive
and a summary of the different possible combinations will be presented at the end of the section.

\subsubsection{Long wavelength approximation}
This approximation is obtained by replacing the second order nonlinear Hamiltonian $H_2$ given by
\eqref{H_2} in the gyrokinetic Lagrangian, \eqref{L_ORB5}, by the nonlinear Hamiltonian model
\cite{Tronko2017} containing FLR expansions up to the second order for both its electrostatic and
electromagnetic parts:
\begin{eqnarray}
\label{H_2FLR}
H_2^{\mathrm{FLR}}=-\frac{m_s c^2}{2B^2}\left|\bm\nabla_{\perp}\phi_1\left({\vec{X}}\right)\right|^2
+\alpha\frac{q_s^2}{2 m_s c^2}\bl A_{1{\|}}(\vec{X})^2+m_s\left(\frac{c}{q_s}\right)^2 \frac{\mu}{B}A_{1\|}\bm\nabla_{\perp} ^2A_{1\|}\left(\vec{X}\right)\br.
\end{eqnarray}
This changes only the term associated with the polarization charge of the quasineutrality equation,
\eqref{quasin_full}, so that \eqref{pol_full} is replaced with
\begin{eqnarray}
  \label{quasin_FLR}
  \Qpol{,LWA} = \epsilon_{\delta}\int \dd \Omega\ f_{\text{eq},s}\ \frac{m_s c^2}{B^2} \bm\nabla_{\perp}\phi_1\cdot\bm\nabla_{\perp}\widehat{\phi}_1,
\end{eqnarray}
for all test functions $\widehat{\phi}_1$. The subscript LWA stands for long wavelength
approximation.  Since the magnetic terms in \eqref{H_2FLR} remain unchanged comparing to the
Hamiltonian $H_2$ given by \eqref{H_2}, as the long wavelength approximation had already been done,
the corresponding Ampère equation remains the same as given by \eqref{GK_Ampere_full}. The
gyrokinetic Vlasov equation is unchanged as well, since the background $H_0$ and linear $H_1$
Hamiltonians are not affected by the approximation and no contributions from the second order
Hamiltonian appear in the characteristics given by \eqref{linear_charakt}.

\subsubsection{Padé approximation}
In addition to the long wavelength approximation, a Padé-approximated quasineutrality model for the
ion species is available in \orb{} \cite{Dominski2017, Lanti2016}. In practice, however, the Padé
approximation is currently only implemented for one ion species ($s=\text{i}$). In order to include
this approximation inside the common variational principle, the linear Hamiltonian model has to be
slightly modified with respect to \eqref{H_1} for both ions:
\begin{equation}
H_{1,\text{Padé}}= \left(1-\bm\nabla_{\perp}\cdot\rho_\text{i}^2\bm\nabla_{\perp}\right) H_1,
\end{equation}
and electrons:
\begin{equation}
H_{1,\text{Padé}}^{\mathrm{dk}}= \left(1-\bm\nabla_{\perp}\cdot\rho_\text{i}^2\bm\nabla_{\perp}\right) H_1^{\mathrm{dk}}.
\end{equation}
The nonlinear Hamiltonian model in that case is given by the FLR second-order truncated Hamiltonian
$H_2^{\mathrm{FLR}}$, \eqref{H_2FLR}. The quasineutrality equation in a weak form is written in a
different way by multiplying it by the operator
$[1-\bm\nabla_{\perp}\cdot\rho_\text{i}^2\bm\nabla_{\perp}]$ to cancel the
$\Big[1-\bm\nabla_{\perp}\cdot\rho_\text{i}^2\bm\nabla_{\perp}\Big]^{-1}$ term in the polarization
density. This is done for computational reasons: the inverse of the block banded matrix coming from
the discretization of the $\Big[1-\bm\nabla_{\perp}\cdot\rho_\text{i}^2\bm\nabla_{\perp}\Big]$
operator is a full matrix. For example, with drift-kinetic electrons, this leads to:
\begin{eqnarray}
  0&=&q_\text{i} \int\dd \Omega\ f_\text{i}\ \left(1-\bm\nabla_{\perp}\cdot\rho_\text{i}^2\bm\nabla_{\perp}\right)\left\langle\widehat{\phi} _1\right\rangle-
  e \int\dd \Omega\ f_\text{e}\ \left(1-\bm\nabla_{\perp}\cdot\rho_\text{i}^2\bm\nabla_{\perp}\right)\ \widehat{\phi}_1
  \nonumber\\
  &+&\epsilon_{\delta}\int \dd \Omega \ f_{\text{eq,i}}\ \frac{m_\text{i} c^2}{B^2} \bm\nabla_{\perp}\phi_1\cdot\bm\nabla_{\perp}\widehat{\phi}_1.
\end{eqnarray}

\subsubsection{\label{adiab}Adiabatic electron model}
In order to include a model with adiabatic electrons inside the variational formulation, we need to
include a fluid approximation for the electron dynamics inside the field-particles Lagrangian. Compared
to the main field-particle Lagrangian, \eqref{L_ORB5}, here the sum over the species in the first
term is over the ion species only and the field term is modified by a purely electrostatic contribution
from the electrons. The action principle for a model with adiabatic electrons is then given by
\begin{align}
\mathcal{A}_{\rm{adiab}}=\int {\mathrm{d}} t\ \mathcal{L}_{\rm{adiab}} =& \sum_{s\neq \text{e}} \int\ \mathrm{d}t\ \mathrm{d}V \mathrm{d}W
\bl\frac{q_s}{c} \vec{A}^*\cdot\dot{\vec{X}}+\frac{m_s c}{q_s}\mu\dot{\theta}-\left(H_0+\epsilon_{\delta} H_1\right)
\br f_s
\nonumber
\\
&+
%\epsilon_{\delta} \int \ \mathrm{d}V \mathrm{d}W\ \left(e F_C \phi_1+\epsilon_{\delta}\frac{e^2}{2 T_e}F_C\left(\phi_1-\overline{\phi}_1\right)^2\right)
\epsilon_{\delta} \int \mathrm{d}t\ \mathrm{d}V \ \bl n_{\text{e}0} \phi_1+\epsilon_{\delta}\ \frac{e}{2 T_\text{e}} n_{\text{e}0}\left(\phi_1-\overline{\phi}_1\right)^2\br
-\epsilon_{\delta}^2\sum_{s\neq \text{e}} \int  \mathrm{d}t\ \mathrm{d}V \mathrm{d}W H_{2} f_{\text{eq},s},
\label{L_ADIAB}
%\vec{Z}=\left(\vec{X},p_{\|},\mu,\theta\right)
\end{align}
where $\overline{\phi}_1$ represents the flux-surface-averaged electric potential given by
\begin{equation}
  \label{eq:FSA}
  \overline{\phi}_1 \equiv \frac{\int \phi_1 \jac(s, \ts) \dd\ts\dd\varphi}{\int \jac(s, \ts) \dd\ts\dd\varphi},
\end{equation}
where $\jac(s, \ts) = \grad s\cdot(\grad\ts\times\grad\varphi)$ is the Jacobian of the magnetic
coordinate transformation and $n_{\text{e} 0}$ is the equilibrium electron density. Since the
adiabatic electron model is only valid in the electrostatic limit, the velocity part of the phase
space volume reduces to $\mathrm{d} W=B_{\|}^*m_s{\mathrm d}v_{\|}{\mathrm d}\mu$ and
${\vec B}^*=\bm\nabla\times{\vec A}^*$ with
$\vec A^*={\vec A}+ (c/q_s) m_s v_{\|}\widehat{\vec{b}}$ while the spatial part remains
unchanged with respect to the electromagnetic case $\mathrm{d}V=\mathrm{d}^3{\vec X}$.  The
Hamiltonian models are now defined for a simplified electrostatic case as
\begin{eqnarray}
\label{H_0adiab}
H_0&=&\frac{m_s v_{\|}^2}{2}+\mu B,\\
\label{H_1adiab}
H_1&=&q_s \left\langle\phi_1\right\rangle.
\end{eqnarray}
The nonlinear ion dynamics is defined by the electrostatic part of either the full FLR, the
Padé-approximated, or the second order FLR long-wavelength-approximated nonlinear Hamiltonian.

The corresponding Vlasov equation does not contain any contribution from the electron species, so we
have for ions ($s=\text{i}$)
\begin{eqnarray}
0=\frac{\dd f_s}{\dd t}=\frac{\partial f_s}{\partial t}
+\dot{\vec X}\cdot\bm\nabla f_s+\dot{v}_{\|}\frac{\partial f_s}{\partial v_{\|}},
\label{eq:Vlasov_ions}
\end{eqnarray}
with the characteristics corresponding to the electrostatic limit ($\alpha=0$) of
Eqs.~\neqref{eq:xdot} and \neqref{eq:pdot}:
\begin{eqnarray}
\label{ion_charakt}
\dot{\vec X}&=&\frac{c\bhat}
{q_s B_{\|}^{*}}\times\bm\nabla\left(\mu B+\epsilon_{\delta} q_s \left\langle\phi_1\right\rangle\right)+\frac{\vec{B}^{*}}{B_{\|}^{*}}v_{\|},\\
\dot{v}_{\|}&=&-\frac{\vec{B}^{*}}{B_{\|}^{*}}\cdot \bm\nabla \left(\mu B+\epsilon_{\delta} q_s \left\langle\phi_1\right\rangle\right).
\nonumber
\end{eqnarray}

For the quasineutrality equation, only the gyro-charge term is modified leading to
\begin{equation}
  \Qpol{,adiab} = \epsilon_{\delta} \int \mathrm{d} V\ \frac{e n_{\text{e}
      0}}{T_\text{e}}\left(\phi_1-\overline{\phi}_1\right)\widehat{\phi}_1 +\int\mathrm{d} V n_{\text{e}0}\ \widehat{\phi}_1.
\end{equation}

\subsubsection{Hybrid electron model}
There is also the possibility to include a hybrid electron model inside the variational
formulation. In that case the fraction of passing electrons designated with a coefficient
$\alpha_\text{P}$ is treated as an adiabatic species, while the fraction of passing electrons is
treated as a drift-kinetic species. At the same time, the ions are treated as kinetic species. The
corresponding action functional is given by
\begin{eqnarray}
%  \label{L_ADIAB}
\mathcal{A}_{\mathrm{hybrid}}=\int\mathrm{d}t\ \mathcal{L}_{\rm{hybrid}}&=& \sum_{s\neq \text{e}} \int\ \mathrm{d} t\  \dd\Omega
\bl\frac{q_s}{c} \vec{A}^*\cdot\dot{\vec{X}}+\frac{m_s c}{q_s}\mu\dot{\theta}-\left(H_0+\epsilon_{\delta} H_1\right)
\br\ f_s
-\epsilon_{\delta}^2\sum_{s\neq \text{e}} \int\ \mathrm{d} t\ \dd\Omega \ H_{2} f_{\text{eq},s}
\nonumber
\\
&+&\int\ \mathrm{d} t\ \mathrm{d} V \int_{\mathrm{trapped}}\mathrm{d}W \bl\frac{e}{c} \vec{A}^*\cdot\dot{\vec{X}}+\frac{m_\text{e} c}{e}\mu\dot{\theta}-\left(H_0+\epsilon_{\delta} H_1^{\mathrm{dk}}\right)\br f_\text{e}
\nonumber
\\
&+&
%\epsilon_{\delta} \int \ \dd\Omega\ \left(e F_C \phi_1+\epsilon_{\delta}\frac{e^2}{2 T_e}F_C\left(\phi_1-\overline{\phi}_1\right)^2\right)
\alpha_\text{P}\ \epsilon_{\delta} \int \mathrm{d} t\  \mathrm{d}V \ \bl n_{\text{e}0} \phi_1+\epsilon_{\delta}\ \frac{e}{2 T_\text{e}} n_{\text{e} 0}\left(\phi_1-\overline{\phi}_1\right)^2\br
,
%\vec{Z}=\left(\vec{X},p_{\|},\mu,\theta\right)
\end{eqnarray}
where the integral over the fraction of trapped electrons in the velocity phase space is assumed
with $\int_{\mathrm{trapped}}\mathrm{d}W$.  The phase space configuration is the same as in the case
of an adiabatic electron model.  The gyrocenter model used for modelling the ion species dynamics is
identical to the one presented for the adiabatic electron model discussed in the previous section,
\ie{} the Hamiltonians $H_0$ and $H_1$ are given by Eqs.~\neqref{H_0adiab}--\neqref{H_1adiab} and
the nonlinear Hamiltonian is coming from either the full FLR, the Padé approximation or the long
wavelength approximation. Concerning the gyrocenter models used for modeling the hybrid electron
dynamics, the equilibrium dynamics is defined with $H_0$ given by \eqref{H_0adiab}. The linear part
of the trapped electron dynamics is defined by the drift-kinetic model defined by \eqref{H_1_FLR}.
%\begin{equation}
%H_1^{\mathrm{dk}}=e \phi_1({\vec X}),
%\end{equation}
%which means that no gyroaveraging effects are taken into the account for electron species.
%The quasineutrality condition is satisfied with writing
%\begin{equation}
%n_i=n_e\Rightarrow \int{\mathrm d} W F=\alpha_P\ n_{e 0}^{\rm{pass}}+\left(1-\alpha_P\right) \int \mathrm{d}W F_{\rm{tr}}
%\end{equation}
The quasineutrality equation is only affected through the gyro-charge term that reads
\begin{equation}
  \label{eq:pol_hyb}
  \Qpol{,hyb}= \alpha_\text{P}\ \epsilon_{\delta} \int \mathrm{d} V\ n_{\text{e} 0} \frac{
    e}{T_\text{e}}\left(\phi_1-\overline{\phi}_1\right)\widehat{\phi}_1 + \int\mathrm{d} V \ n_{\text{e} 0 }\ \widehat{\phi}_1+ \int_{\mathrm{trapped}} \dd \Omega\ f_\text{e} \widehat{\phi}_1.
\end{equation}

The ion characteristics are reconstructed identically to the case with adiabatic electrons,
accordingly to \eqref{ion_charakt}. The characteristics for the electrons are defined by
the simplified drift-kinetic equations corresponding to the dynamics of $H=H_0+H_1^{\rm FLR}$:
\begin{eqnarray}
\dot{\vec X}&=&\frac{c\bhat}
{q_s B_{\|}^{*}}\times\bm\nabla\left(\mu B+\epsilon_{\delta} e \ \phi_1\right)+\frac{\vec{B}^{*}}{B_{\|}^{*}}v_{\|},\\
\dot{v}_{\|}&=&-\frac{\vec{B}^{*}}{B_{\|}^{*}}\cdot \bm\nabla \left(\mu B+\epsilon_{\delta} e\ \phi_1\right).
\nonumber
\end{eqnarray}

The hybrid electron model presented above was originally implemented to simulate linear electron
modes such as TEM with a larger timestep than with fully drift-kinetic electrons. However, in
nonlinear regime, it does not ensure the ambipolar condition---which also impacts the conservation
of the toroidal angular momentum---as no flux-surface-averaged passing-electron density is accounted
for. Furthermore, due to trapping/de-trapping processes, the former hybrid electron model adds
spurious sources of \eg{} particles and momentum. To address this issue, an upgraded hybrid electron
model has been implemented in \orb{} as an improvement of the model presented in
\cite{Idomura2016}. The idea of this updated model is to take into consideration only the $n=m=0$
component of the passing electron density while keeping an adiabatic response for the other passing
components. This way, the quasineutrality equation, \eqref{eq:pol_hyb}, is slightly changed as follows
\begin{equation}
  \Qpol{,hyb}= \alpha_\text{P}\ \epsilon_{\delta} \int \mathrm{d} V\ n_{\text{e} 0} \frac{
    e}{T_\text{e}}\left(\phi_1-\overline{\phi}_1\right)\widehat{\phi}_1 + \int\mathrm{d} V \ n_{\text{e} 0 }\ \widehat{\phi}_1+ \int_{\mathrm{trapped}} \dd \Omega\ f_\text{e} \widehat{\phi}_1+ \int_{\mathrm{passing}} \dd \Omega\ f_\text{e} \widehat{\phi}_1^{00},
\end{equation}
where $\widehat{\phi}_1^{00}$ is the $n=m=0$ component of the arbitrary test function $\widehat{\phi}_1$.

\subsubsection{Summary of the models}
\label{sec:summary}
All the variants of the particle models presented in the previous sections are summarized here. The
main changes brought by the different models mainly come through the quasineutrality equation which
can be written
\begin{align}
  \label{quasin_full_gen}
  &\sum_{s\neq\text{e}}\Qgyr{}+\Qdk{}= \sum_{s\neq\text{e}} \Qpol{},\\
  &\Qgyr{} = \int \dd\Omega\ f_s\ q_s\left\langle\widehat{\phi}_1\right\rangle,\\
  &\Qdk{} = -\int \dd\Omega\ f_\text{e}\ e \,\widehat{\phi}_1(\vec{X}),\\
  &\Qpol{} = \epsilon_{\delta}\int \dd\Omega\ f_{\text{eq},s}\ \frac{q_s^2}{B}\frac{\partial}{\partial\mu}\left(\left\langle\phi_1\widehat{\phi}_1\right\rangle-\Big\langle\phi_1\Big\rangle\left\langle\widehat{\phi}_1\right\rangle\right),
%  &\sum_{s}\int \dd\Omega\ F\ q_s\left\langle\widehat{\phi}_1\right\rangle
%  =\epsilon_{\delta} \sum_{s}\int \dd\Omega\ F_C\ \frac{q_s^2}{B m_s}\frac{\partial}{\partial\mu}\left(\left\langle\phi_1\widehat{\phi}_1\right\rangle-\left\langle\phi_1\right\rangle\left\langle\widehat{\phi}_1\right\rangle\right),
\end{align}
%\begin{align}
  %&\sum_{s} \Qgyr{} = \sum_{s} \Qpol{},\\
  %&\Qgyr{} = \int \dd\Omega\ F\ q_s\left\langle\widehat{\phi}_1\right\rangle,\\
  %&\Qpol{} = \epsilon_{\delta} \sum_{s}\int \dd\Omega\ f_{eq}\ \frac{q_s^2}{B}\frac{\partial}{\partial\mu}\left(\left\langle\phi_1\widehat{\phi}_1\right%\rangle-\left\langle\phi_1\right\rangle\left\langle\widehat{\phi}_1\right\rangle\right),
%\end{align}
where $\Qgyr{}$ is the term corresponding to the ion gyrodensity contribution, $\Qdk{}$ is the term
corresponding to the electron drift kinetic density contribution and $\Qpol{}$ represents the
polarization density contribution from the ions. For the ions, only $\Qpol{}$ is affected by the
different models:
\begin{align}
  \label{eq:sumions}
  \Qpol{,LWA} &= \epsilon_{\delta}\int \dd\Omega\ f_{\text{eq},s}\ \frac{m_s c^2}{B^2} \bm\nabla_{\perp}\phi_1\cdot\bm\nabla_{\perp}\widehat{\phi}_1,\\
  \mathcal{Q}_{i,\text{Padé}}^{\text{pol}} &= \epsilon_{\delta}\int \dd\Omega\ f_{\text{eq,i}}\ \frac{m_\text{i} c^2}{B^2} \Big[1-\bm\nabla_{\perp}\cdot\rho_\text{i}^2\bm\nabla_{\perp}\Big]^{-1}\Big[\bm\nabla_{\perp}\phi_1\cdot\bm\nabla_{\perp}\widehat{\phi}_1\Big].
\end{align}
Note that in the case of the Padé approximation, all the quasineutrality equation is multiplied by
$\Big[1-\bm\nabla_{\perp}\cdot\rho_\text{i}^2\bm\nabla_{\perp}\Big]$ to avoid inverting it.

For the electrons, only $\Qdk{}$ is changed by the different fluid and hybrid approximations:
\begin{align}
  \label{eq:sumelectrons}
  \Qdk{,adiab} &= \epsilon_{\delta} \int \mathrm{d} V\ \frac{e n_{\text{e}
                 0}}{T_\text{e}}\left(\phi_1-\overline{\phi}_1\right)\widehat{\phi}_1 -\int\mathrm{d} V n_{\text{e} 0 }\ \widehat{\phi}_1,\\
  \Qdk{,hyb} &=\alpha_\text{P}\ \epsilon_{\delta} \int \mathrm{d} V\ \frac{
               e n_{\text{e} 0} }{T_\text{e}}\left(\phi_1-\overline{\phi}_1\right)\widehat{\phi}_1 - \int\mathrm{d} V \ n_{\text{e}
               0 }\ \widehat{\phi}_1- e\int_{\mathrm{trapped}} \dd \Omega f_\text{e} \widehat{\phi}_1+ \int_{\mathrm{passing}} \dd \Omega\ f_\text{e} \widehat{\phi}_1^{00}.
\end{align}
In \orb{}, the previous approximations are not mutually exclusive, \ie{} each model for the
polarization density can be combined with any electron model.

\subsection{$\delta f$ method and background distribution functions}
\label{sec:distrib}
The \orb{} code uses a $\delta f$ control-variate approach to reduce the numerical noise due to the
finite phase-space sampling \cite{Aydemir1994a,Hatzky2019}. The rationale of this method is to
separate the total distribution function into two parts: a time-independent part $f_0$ and a
time-dependent part $\delta f$. The first function, $f_0$, is supposed to be known and easily
computable. Only the $\delta f$ part is represented with a sample of ``numerical particles'' or
``markers''. The statistical sampling error will thus be reduced, as compared to a full-$f$ method,
if $|\delta f| \ll f_0$.

In the collisionless limit and in the absence of sources, the total distribution function is
conserved along the trajectories. Using the $\delta f$ separation, we obtain
\begin{equation}
  \label{eq:fulldt}
  \frac{\dd \delta f}{\dd t} = -\left.\frac{\dd f_0}{\dd t}\right|_0-\left.\frac{\dd f_0}{\dd t}\right|_1,
\end{equation}
where the time-derivative operator has been split into two parts labeled by 0 and 1 and which
respectively represent the unperturbed dynamics, \ie{} without the fluctuating fields, and the
perturbed. In the \textit{standard $\delta f$} method, we choose $f_0\equiv f_{\text{eq}}$ to be an
equilibrium distribution, solution of the unperturbed collisionless equations of motion and thus, to
satisfy $\{f_{\text{eq}}, H_0\}=0$ reducing \eqref{eq:fulldt} to
\begin{equation}
  \label{eq:eqdt}
  \frac{\dd \delta f}{\dd t} = -\left.\frac{\dd f_{\text{eq}}}{\dd t}\right|_1.
\end{equation}

In \orb{}, different choices for the initial distribution function are available. The plasma can be
supposed to be in a local thermodynamic equilibrium described by a \textit{local Maxwellian}
$f_\text{L}(\psi, \epsilon, \mu)$. Both the particle energy $\epsilon$ and the magnetic moment $\mu$
are constants of motion but the poloidal magnetic flux $\psi$ is not. The local Maxwellian is
therefore not invariant under the unperturbed dynamics and \eqref{eq:fulldt} must be used. The
inclusion of the $\left.\dd f_\text{L}/\dd t \right|_0$ term leads to the drive of a spurious zonal
flow discussed in \cite{Idomura2003} which appears already in the linear phase of a simulation even
though zonal flows are linearly stable and excited through nonlinear coupling \cite{Lin1998a}. As
done in many PIC codes, the term responsible for this zonal flow drive can be neglected but it is
not consistent with the perturbative ordering used here.

The other approach is to use a distribution function that is a true equilibrium, \ie{} that $f_0$ is
a function of constants of motion only. This is the so-called \textit{canonical} Maxwellian
$f_{\text{C}}(\psi_0, \epsilon, \mu)$, where $\psi_0=\psi + (m_sc/q_s)(F(\psi)/B) v_\|$ is the
toroidal momentum which is conserved in an axisymmetric toroidal system. However, it is easily shown
that the effective density and temperature computed from $f_\text{C}$ are different from the ones
given as input and function of $\psi$. The use of a canonical Maxwellian can lead to large,
unrealistic values of parallel flows preventing any instability to develop, especially for small
system size and large $n_0$ and $v_\|$ gradients \cite{Angelino2006}.

To address this issue, a \textit{corrected canonical Maxwellian}
$f_{\text{CC}}(\hat{\psi}, \epsilon, \mu)$ is used. A correction term is added to the toroidal
momentum to minimize the gap between the local and canonical Maxwellians while still being an
equilibrium distribution. The corrected toroidal momentum reads
\begin{equation}
  \label{eq:corrterm}
  \hat{\psi} = \psi_0 + \psi_{0,\text{corr}} = \psi_0 - \sign(v_\|)\frac{m_sc}{q_s}R_0\sqrt{2(\epsilon-\mu B_0)}\mathcal{H}\pl\epsilon-\mu B_0 \pr,
\end{equation}
where $R_0$ is the major radius and $\mathcal{H}$ is the Heaviside function. The correction term is
zero for trapped particles and of opposite sign for forward and backward passing particles. The
corrected toroidal momentum being built only with constants of motion,
$f_{\text{CC}}(\hat{\psi}, \epsilon, \mu)$ satisfies $\{f_{\text{CC}}, H_0\}=0$.

\subsection{Strong flows}
The strong flow gyrokinetic ordering allows for $u_E/v_{\text{th,i}} \sim 1$, with
$\vec{u}_E=c\pl\bhat\times\bm\nabla\Phi/B\pr$ the background E$\times$B velocity, where $\Phi$
represents the background electric potential, and $v_{\text{th,i}}=\sqrt{T_\text{i}/m_\text{i}}$ is
the ion thermal velocity \cite{Hahm1996}. Implementing this ordering in \orb{} enables the treatment
of plasmas rotating toroidally at close to the Mach velocity. More details of this formalism have
been published earlier \cite{Collier2016}.  In that case, a further approximation is performed on
the background distribution function. While a local Maxwellian is used for the polarization density
in the quasineutrality equation, the canonical Maxwellian is implemented for the reconstruction of
the gyrokinetic Vlasov equation.

In order to include the model containing a background electrostatic potential $\Phi$ within the
general field-gyrocenter action given by \eqref{L_ORB5}, the background Hamiltonian $H_0$ as well
as the symplectic magnetic potential $\vec{A}^*$ have to be consistently modified:
\begin{equation}
H_{0}^{\rm{flow}}= q_s\Phi+\mu B+\frac{p_{z}^2+(m_s\vec{u}_E)^2}{2 m_s},
\label{H_0flow}
\end{equation}
and ${\vec A}^*={\vec A}+(c/q_s)\ p_{z}\bhat+(m_s c/q_s){\vec u}_E$.

The field part of the model with a background $E\times B$ velocity is assumed to be in the
electrostatic limit. This corresponds to setting $\alpha=0$ in \eqref{L_ORB5}. Remark that including
the background $E\times B$ velocity does not affect the quasineutrality equation, since no
corrections due to the presence of a strong flow are included into the linear and nonlinear
Hamiltonian models given by Eqs.~\neqref{H_1}-\neqref{H_2}. The gyrokinetic Vlasov equation is
modified according to the change of background dynamics from the $H_0$ given by \eqref{H_0} to
$H_{0}^{\rm{flow}}$ given by \eqref{H_0flow}.  The corresponding $\delta f$ gyrokinetic Vlasov
equation is reconstructed from the modified characteristics.
%, taking into the account changes of background dynamics.
Since the perturbed magnetic field is not considered, $p_{z}=m_s v_{\|}$ is a purely kinetic momentum:
\begin{eqnarray}
\dot{\vec X}&=&\frac{c\bhat}
{q_s B_{\|}^{*}}\times\bm\nabla\left(q_s \Phi + \mu B+\frac{m_s}{2}|\vec{u}_E|^2+\epsilon_{\delta} q_s \left\langle\phi_1\right\rangle\right)+\frac{\vec{B}^{*}}{B_{\|}^{*}}\frac{p_{z}}{m_s},\\
\dot{p}_{z}&=&-\frac{\vec{B}^{*}}{B_{\|}^{*}}\cdot \bm\nabla \left(q_s \Phi+\mu B+ \frac{m_s}{2}|\vec{u}_E|^2+\epsilon_{\delta} q_s \left\langle\phi_1\right\rangle\right).
\nonumber
\end{eqnarray}

For strongly rotating plasmas, with Mach number around one, dynamic pressures due to the flow are
comparable to the thermal pressure, and a modified Grad-Shafranov equation should be used to
accurately compute the magnetic equilibrium. To self-consistently include these effects, we have
used the MHD code FLOW \cite{Guazzotto2004} which can solve the MHD force balance equation in the
presence of a background flow. FLOW reads the equilibrium via the standard EQDSK format
\cite{Lutjens1996}. We have considered only toroidally rotating MHD equilibria, with the temperature
being a flux surface function, as this allows collisionless kinetic and MHD equilibria to be
consistent in the large-system size limit.

\subsubsection{Global gyrokinetic equilibria for rotating plasmas}

The constants of motion are the magnetic moment, $\mu$, the unperturbed energy of the particle,
$\varepsilon = H_{0}^{\rm{flow}}$, the sign of the parallel velocity (for passing particles), and
finally the toroidal canonical momentum, $\psi_0$, which is conserved in an tokamak due to
axisymmetry. The strong-flow canonical momentum $\psi_\text{C}$ is an extension of the canonical
momentum in the presence of strong flows:
\begin{equation}
  \psi_\text{C} = \psi + \frac{m_sc}{q_s}\frac{F}{B}{v}_{\parallel} + \frac{m_sc}{q_s}u_{\varphi},
\end{equation}
where $u_{\varphi}$ is the toroidal component of the background $E\times B$ velocity.

In the presence of toroidal rotation, the canonical Maxwellian, which is corrected so that the flux
surface averaged density remains close to $n_0$ when rotation is introduced, is given by
\begin{equation}
  \label{dist_corrected}
  f_\text{C}= \left(\frac{m_s}{2\pi{T_0(\psi_\text{C})}}\right)^{3/2}n_0(\psi_\text{C}) \exp
  \curl -\frac{1}{T_0(\psi_\text{C})} \bl H_{0}^{\rm{flow}} + \frac{m_s R_0(\psi_\text{C})^2}{2} \pl\frac{\partial\overline{\Phi}}{\partial\psi}\pr^2 \br \curr,
\end{equation}
where $\overline{\Phi}$ is the flux surface average of $\Phi$. In the local limit, this choice leads
to a in-out density variation
\begin{equation}
n_s = n_0(\psi)\exp\bl \frac{m_s (R^2 - R_0^2) \Omega^2}{2}\br,
\end{equation}
where the plasma rotation frequency $\Omega$ may be expressed as $\Omega = \partial\Phi/\partial\psi$.

\subsection{Collisions}
\label{sec:collisions}
The inclusion of collisions in a gyrokinetic code like \orb{} is important to assess the right level
of transport. Indeed, collisions are required to model the neoclassical physics, which is a key
player in the transport of certain classes of particles, \eg{} heavy impurities. Furthermore,
collisions are known to impact turbulence. For example, ITG driven turbulence increases when
collisions are taken into account due to the collisional damping of the zonal flows
\cite{Dif-Pradalier2009,Vernay2012}. On the other hand, TEM turbulence is reduced by collisions via
the collisional detrapping of electrons.

\orb{} currently includes ion-ion intraspecies and electron-ion collisions \cite{Vernay2010}. For
the collisional dynamics, FLR effects are neglected. In \orb{}, collisions are represented by a
linearized Landau collision operator. The linearization procedure is done with respect to a local
Maxwellian background $f_\text{L}$ which is in the kernel of the full collision operator. The full
Landau operator describing the effect of the distribution function $f$ on itself may be decomposed
into four terms:
$C_{\text{ab}}[f_\text{b}, f_\text{a}] = C_{\text{ab}}[f_{\text{b,L}}, f_{\text{a,L}}] +
C_{\text{ab}}[f_{\text{b,L}}, \delta f_\text{a}] + C_{\text{ab}}[\delta f_\text{b}, f_{\text{a,L}}]
+C_{\text{ab}}[\delta f_\text{b}, \delta f_\text{a}]$, where $\delta f_s$ is the perturbed part of
the distribution of the species $s$. Note that in our notation, $C[f_\text{a}, f_\text{b}]$ refers
to the effect of $f_\text{a}$ on $f_\text{b}$. Note that, for the whole collision part, the species
background distribution function is converted to a local Maxwellian if it is not already the
case. After, the collision dynamics has been treated the background Maxwellian is converted back to
its original form if needed. For Maxwellian distributions with identical parallel
velocities and temperatures, the first term on the right-hand side is zero. Assuming the
perturbation is small, the final, nonlinear term, is also neglected leaving two terms called the
``test particle'' term $C_{\text{ab}}[f_{\text{b,L}}, \delta f_\text{a}]$ and the ``background
reaction'' term $C_{\text{ab}}[\delta f_\text{b}, f_{\text{a,L}}]$.

For the self-collisions, the ``test particle'' term can be readily evaluated using the exact Landau
operator in its drag-diffusion form:
\begin{equation}
  \label{eq:coll1}
  C[f_{\text{L}}, \delta f] = \frac{\partial}{\partial \vec{v}} \cdot \bl \vec{\Gamma} (f_{\text{L}}) \delta f \br - \frac{\partial^2}{\partial \vec{v}  \vec{v}} : \bl\overline{\overline{D}} (f_{\text{L}}) \delta f \br,
\end{equation}
where the drag vector and the diffusion tensor are respectively given by
\begin{equation}
  \label{eq:coll2}
  \vec{\Gamma} = - \bar{\nu} H (x) \vec{v},
  \quad
  \overline{\overline{D}} = \frac{\bar{\nu} v_{\text{th}}^2}{4} \bl K(x) \pl \overline{\overline{I}} - \frac{\vec{v} : \vec{v}}{v^2} \pr + 2 H(x) \frac{\vec{v} : \vec{v}}{v^2} \br,
\end{equation}
where the collision frequency is defined as
$\bar{\nu} = 8\pi n q^4 \ln \Lambda / m^2 v_{\text{th}}^3$, $x=v / \sqrt 2 v_{\text{th}}$ is the
normalized velocity with $v_{\text{th}} = \sqrt{T/m}$ the thermal velocity of the species, and
$\overline{\overline{I}}$ is the identity tensor. The Coulomb logarithm, $\ln \Lambda$, is assumed
constant across the plasma, and is typically having a value of 10--15. The functions $K(x)$ and
$H(x)$ are resulting from the analytical evaluation of the Rosenbluth potentials of Maxwellian
distributions:
\begin{align}
  \label{eq:coll3}
  H(x) &= \frac{1}{2\sqrt 2 x^3} \bl\erf(x) - x\erf'(x)\br,\\
  \label{eq:coll4}
  K(x) &= \frac{1}{\sqrt 2 x} \phi(x) - H(v).
\end{align}
where $\erf$ represents the error function.

Evaluating the background reaction term exactly would require the reconstruction of the $\delta f$
distribution function and the evaluation of integrals over velocity space. Such a direct approach is
too expensive and includes steps subject to significant noise in a PIC code. Instead, \orb{} uses an
approximation first suggested by \cite{Lin1995a}:
$C[\delta f, f_{\text{L}}] \simeq f_{\text{L}} \beta (\delta f)$, with
\begin{equation}
  \label{eq:coll5}
  \beta(\delta f) = \frac{1}{n} \bl 6 \sqrt{\pi}  H(x) \frac{\delta P_{||} v_{||}}{v_{\text{th}}^2}  + \sqrt{ \pi} G(x) \frac{\delta E}{v_{\text{th}}^2} \br,
\end{equation}
where $G(v) = (4x^2 - 1) H(x) - K(x)$. The two terms $\delta P_{||}$ and $\delta E$ represent
respectively the parallel momentum and energy transferred to the distribution by the ``test
particle'' operator. This approximation can be shown to satisfy the desirable properties of a
collision operator \cite{Lin1995a, Brunner1999}. Indeed, it conserves the mass and, when combined
with its counterpart $C[f_{\text{L}}, \delta f]$, conserves also the momentum and
energy. Furthermore, the combined linear operator is self-adjoint and satisfies the H-theorem. The
operator is zero if the perturbation is a shifted linearized Maxwellian, \ie{} such distributions
are stationary states.

The only interspecies collisions which are currently taken into account in \orb{} are the
electron-ion collisions. The ``test particle'' part of the electron-ion collisions in \orb{} is
represented by a Lorentz operator, which assumes a large mass ratio between ions and electrons. In
this limit, electrons experience only pitch-angle scattering. This Lorentz operator can simply be
written:
\begin{equation}
  \label{eq:coll6}
  C_{\text{ei}}[f_\text{i},\delta f_\text{e}] = - \nu_{\text{ei}}\pl v \pr\frac{\partial}{\partial \xi} \bl (1 - \xi^2) \frac{\partial \delta f_\text{e}}{\partial \xi} \br,
\end{equation}
where the electron-ion collision frequency is given by
$\nu_{\text{ei}}\left(v \right)=(\bar{\nu}_{\text{ei}} / 4) (v_{\text{th,e}} / v)^3$, with
$\bar{\nu}_{\text{ei}} = 8\pi n_\text{i} Z^2 e^4 \ln \Lambda / m_\text{e}^2 v_{\text{th,e}}^3$ and
$\xi$ is the pitch angle. The ``test particle'' Lorentz operator conserves the mass and energy. The
``background reaction'' of the Lorentz operator is neglected in \orb{}. Therefore, momentum
conservation is not ensured by the reduced electron-ion collision operator.

\subsection{Conservation laws and diagnostics}
In this section we present the conserved quantities associated with the field-particle Lagrangian,
which are implemented in \orb{} as diagnostic tools. These quantities can be obtained from a direct
application of the Noether method, details of the derivation can be found in \cite{Tronko2016}. We
start with presenting the energy invariant corresponding to each model. This invariant is used for
constructing the so-called power balance diagnostics, which allows one to verify the quality of
numerical simulations.

The power balance diagnostic is naturally included in the Lagrangian framework. Indeed, it can be
directly and exactly obtained from the energy conservation law, which is related to the Lagrangian
of the given physical system via the Noether method. In this section we give expressions of the power
balance diagnostics corresponding to the models implemented in the \orb{} code.

First, we provide a generic expression for the energy density corresponding to the most complete
electromagnetic model, which can also be obtained from a direct application of the Noether method, see
\eg{} \cite{Tronko2016}.
\begin{eqnarray}
\label{E_em}
\mathcal{E}^{\mathrm{EM}}&=& \sum_{s} \int  \dd \Omega\  H_0\ f_s+
\epsilon_{\delta}\sum_{s\neq\text{e}} \int  \dd \Omega\  H_1\ f_s
+\epsilon_{\delta} \int  \dd \Omega\  H_1^{\mathrm{dk}}\ f_\text{e}
\\
\nonumber
&+&\epsilon_{\delta}^2 \sum_{s\neq\text{e}} \int \dd \Omega\  H_2\ f_{\text{eq},s} +
\alpha\epsilon_{\delta}^2 \int \dd \Omega\  H_2^{\mathrm{dk}}\ f_{\text{eq,e}}+ \alpha\
\epsilon_{\delta}^2\int \dd V \frac{\left|\bm\nabla_{\perp} A_{1\|}\right|^2}{8\pi}.
\end{eqnarray}
This expression can be simplified and rewritten in the form of code diagnostics by direct
substitution of the expression for the Hamiltonians $H_0$, $H_1$ given by Eqs.~\neqref{H_0} and
\neqref{H_1}, while $H_2$ is given by \eqref{H_2} in the case of the all-orders polarization density
model and by \eqref{H_2FLR} in the case of the long-wavelength approximation. At the next step, the
second term in the expression for energy is rewritten using the corresponding quasineutrality and
Amp\`ere equations in their weak form. Here we choose a particular test function
$\widehat{\phi}_1=\phi_1$ and we substitute it in Eqs.~\neqref{quasin_full}--\neqref{pol_full} or,
for the case of the long-wavelength approximation, in Eqs.~\neqref{quasin_full}, \neqref{gyr_full},
\neqref{dk_full}, and \neqref{quasin_FLR}. Similarly, the test function $\widehat{A}_{1\|}=A_{1\|}$
is substituted to the corresponding Amp\`ere equation given by \eqref{GK_Ampere_full}. In PIC codes
particles and fields are evaluated in two different ways: particles are advanced continuously along
their characteristics while fields are evaluated on a fixed grid. To control the quality of the
simulation, the contributions to the energy from the particles and from the fields should be
computed independently. This is why we are considering the power balance equation, also called the
$E\times B$ transfer equation. The code diagnostics is implemented to verify the following balance
equation for $\mathcal{E}^{\mathrm{EM}}=\mathcal{E}_\text{F}+\mathcal{E}_{\mathrm{kin}}$:
\begin{equation}
0=\frac{{\mathrm d}{\mathcal E}^{\mathrm{EM}}}{{\mathrm d} t}\Rightarrow\frac{{\mathrm d}{\mathcal E}_{\mathrm{kin}}}{{\mathrm d} t}=-\frac{{\mathrm d}{\mathcal E}_{\mathrm{F}}}{{\mathrm d} t},
\end{equation}
where the time derivative of the l.h.s. can be evaluated through the particles characteristics and
the r.h.s. from the fields contributions evaluated on the grid.

From \eqref{E_em}, the first term on the r.h.s. is defined as the ``kinetic energy''
$\mathcal{E}_{\text{kin}}$:
\begin{equation}
  \mathcal{E}_{\text{kin}} = \sum_{s} \int  \dd \Omega\  H_0\ f_s = \sum_{s}\int \dd\Omega\ \left(\frac{p_{z}^2}{2 m_s}+\mu B\right) f_s,
\end{equation}
which depends only on the unperturbed Hamiltonian $H_0$ and therefore, its time derivative can be
evaluated considering only the unperturbed characteristics. The other terms are defined as the
``field energy'' $\mathcal{E}_{\text{F}}$, which can be written, for the case of the $H_2$
Hamiltonian written in the LWA, \eqref{H_2FLR}, as:
\begin{align}
  \mathcal{E}_{\text{F}} = & \epsilon_{\delta} \sum_{s\neq\text{e}}\int \dd\Omega
                             q_s\left\langle\phi_1-\alpha A_{1\|}\frac{p_{z}}{m_s} \right\rangle\
                             f_s -\epsilon_{\delta}\int \dd\Omega\, e\left(\phi_1-\alpha A_{1\|}\frac{p_{z}}{m_\text{e}} \right)\ f_\text{e}\\
                           & +\epsilon_{\delta}^2\sum_{s\neq\text{e}}\int \dd\Omega f_{\text{eq},s} \curl -\frac{m_s c^2}{2 B^2} \left|{\bm\nabla}_{\perp}\phi_1\right|^2+
                             \alpha\frac{q_s^2}{2 m_s}\bl A_{1\|}^2+\left(\frac{m_s}{q_s}\right)^2\frac{\mu}{B}
                             A_{1\|}\bm{\nabla}_{\perp}^2 A_{1\|}\br\curr \\
                           & +\epsilon_{\delta}^2\alpha\int \dd\Omega\ f_{\text{eq,e}}\frac{e^2}{2 m_\text{e}}\ A_{1\|}^2 + \epsilon_{\delta}^2\alpha\int \dd V \frac{\left|\bm\nabla_{\perp} A_{1\|}\right|^2}{8\pi}.
\end{align}
Using the quasineutrality equation Eqs.~\neqref{quasin_full}--\neqref{dk_full} with the polarization term in the LWA,
\eqref{quasin_FLR}, Ampère equation, \eqref{GK_Ampere_full}, and setting $\widehat{\phi}_1 = \phi_1$
and $\widehat{A}_{1\|} = A_{1\|}$, we obtain two equivalent expressions for the field energy:

\begin{equation}
\label{E_F_density}
  \mathcal{E}_\text{F} = \epsilon_{\delta}\frac{1}{2}\sum_{s\neq\text{e}}q_s\int \dd\Omega\ \left(\left\langle\phi_1\right\rangle-\alpha\frac{p_{z}}{m_s}\left\langle A_{1\|}\right\rangle\right) f_s
                           -\epsilon_{\delta}\frac{1}{2}e\int \dd\Omega\ \left(\phi_1-\alpha\frac{p_{z}}{m_\text{e}}
                           A_{1\|}\right)\ f_\text{e}.
\end{equation}
Note that \eqref{E_F_density} does not depend on the particular choice for the nonlinear Hamiltonian
$H_2$. Indeed, \eqref{E_F_density} is also valid for the all order FLR polarization density,
\eqref{pol_full}. This is a direct consequence of the fact that the equations of motion, which are
used for rewriting the expression of the energy are obtained from the same field-particle
Lagrangian.

Similarly, a second expression for the field energy written in terms of the polarizations and
magnetizations and depending on the expression of the nonlinear Hamiltonian $H_2$ can be
obtained. For the full FLR polarization density given by \eqref{pol_full}, the alternative field
energy is given by
\begin{align}
\mathcal{E}_\text{F}=&
\frac{1}{2}\sum_{s\neq\text{e}}\epsilon_{\delta}\int \dd\Omega\ f_{\text{eq},s}\ \frac{q_s^2}{B}\frac{\partial}{\partial\mu}\left\langle\widetilde{\phi}_1\left({\vec{X}+
\bm\rho_0}\right)^2\right\rangle
+
\frac{1}{2}\ \alpha\sum_{s\neq \text{e}}\epsilon_{\delta}\int \dd\Omega f_{\text{eq},s}\ \left(\frac{q_s^2}{m_s}A_{1\|}^2
+\frac{\mu}{B}
A_{1\|}\bm\nabla_{\perp}^2 A_{1\|}
\right)\\
&+\frac{1}{2}\ \alpha\ \epsilon_{\delta}\int \dd\Omega f_{\text{eq,e}}\ \frac{e^2}{m_\text{e}}A_{1\|}^2
+\alpha\ \epsilon_{\delta}\int \frac{\mathrm{d}V}{8\pi}\
\left|\bm\nabla_{\perp} A_{1\|}\right|^2.
\end{align}
For the polarization density in the LWA, \eqref{quasin_FLR}, the field energy becomes
\begin{align}
  \label{E_F_fields}
  \mathcal{E}_\text{F}^{\text{LWA}} =& \frac{1}{2}\sum_{s\neq\text{e}}\epsilon_{\delta}\int \dd\Omega\ \frac{m_s c^2}{B^2}\ f_{\text{eq},s}\ \left|\bm\nabla_{\perp}\phi_1\right|^2
                          +
                          \frac{1}{2}\alpha\sum_{s\neq \text{e}}\epsilon_{\delta}\int \dd\Omega f_{\text{eq},s}\ \left(\frac{q_s^2}{m_s c^2}A_{1\|}^2
                          +\frac{\mu}{B}
                          A_{1\|}\bm\nabla_{\perp}^2 A_{1\|}
                          \right) \\
                        &+\frac{1}{2}\alpha\epsilon_{\delta}\int \dd\Omega f_{\text{eq,e}}\
                            \frac{e^2}{m_\text{e} c^2}A_{1\|}^2
                            +\alpha\epsilon_{\delta}\int \frac{\mathrm{d}V}{8\pi}\
                            \left|\bm\nabla_{\perp} A_{1\|}\right|^2.
\end{align}
For the Padé approximated model, the expression for the field energy is
\begin{align}
\mathcal{E}_\text{F}^{\text{Padé}}=&\epsilon_{\delta}\frac{1}{2}q_\text{i}\int \dd\Omega\ \left(1-\bm\nabla_{\perp}\cdot\rho_\text{i}^2\bm\nabla_{\perp}\right)\left\langle\phi_1\right\rangle f_\text{i} -
\epsilon_{\delta}\frac{1}{2}e\int \dd\Omega\ \left(1-\bm\nabla_{\perp}\cdot\rho_\text{i}^2\bm\nabla_{\perp}\right) \phi_1\ f_\text{e}
\\
&+
\epsilon_{\delta}\sum_{s}\int \dd\Omega\ f_s\left(\frac{p_{z}^2}{2 m_s}+\mu B\right).
\nonumber
\end{align}

In the case of the model with adiabatic and hybrid electrons, the expressions for the conserved energy
have to be discussed separately since they are issued from a slightly different variational
formulation, which combines a fluid and kinetic formalism. With adiabatic electrons, the
corresponding contribution to the energy should be considered as a field term:
\begin{eqnarray}
\mathcal{E}= \sum_{s\neq\text{e}} \int \dd\Omega  \left(H_0+\epsilon_{\delta} H_1\right) \ f_s + \epsilon_{\delta}^2\sum_{s\neq\text{e}} \int \dd\Omega  H_2\ f_{\text{eq},s} +
\epsilon_{\delta} \int \ \mathrm{d}V \ \bl n_{\text{e} 0} \phi_1+\epsilon_{\delta}\ \frac{e}{2 T_\text{e}} n_{\text{e} 0}\left(\phi_1-\overline{\phi}_1\right)^2\br,
\end{eqnarray}
where the last term is considered as a field term that includes the energy of the adiabatic
electrons in the system. Following the general procedure, we substitute the test function
$\widehat{\phi}_1=\phi_1$ into the quasineutrality equation. The field energy is then given by
\begin{equation}
\mathcal{E}_\text{F}=
\epsilon_{\delta}\frac{1}{2} \sum_{s\neq\text{e}}\int \dd\Omega q_s \ f_s\ \left\langle\phi_1\right\rangle+\epsilon_{\delta}\frac{1}{2} \int {\mathrm d}V \ n_{\text{e} 0}\ \phi_1
\end{equation}
and the kinetic part of energy consists of the ion contribution only:
\begin{equation}
{\mathcal E}_{\mathrm{kin}}=\sum_{s\neq\text{e}}\int \dd\Omega\ f_s \ H_0=
\sum_{s\neq\text{e}}\int \dd\Omega\ f_s\left(\frac{m_s v_{\|}^2}{2 }+\mu B\right).
\end{equation}

%%% Local Variables:
%%% mode: latex
%%% TeX-master: "../ORB5Review"
%%% End:

%% file: sections/numerical_implementation.tex
\section{Numerical implementation}
\label{sec:numericalimplementation}
\orb{} uses a low-noise $\delta f$ PIC method \cite{Parker1993,Hu1994} consisting of separating the
full distribution function $f$ into a prescribed, time-independent background distribution $f_0$ and
a perturbed, time-dependent distribution $\delta f$ such that only the latter is discretized using
markers, or numerical particles, that are used to sample the phase space. Furthermore, the code uses
a operator splitting approach which consists of solving first for the collisionless dynamics and then
considering the collisions and various sources. The time integration of the collisionless dynamics
is made using a $4$th-order Runge-Kutta (RK4). The collisions are treated with a Langevin approach.

This section describes the numerical implementation of the gyrokinetic equations presented in the
previous section. First, the low-noise $\delta f$ PIC method as well as the field discretization and
solving are presented. Then, the noise reduction techniques, essential to control the unavoidable
noise inherent to the finite sampling of the phase space, are described. Finally, the different heat
sources, relevant diagnostics, and the parallelization of the code are discussed. In this section,
we omit the subscripts $s$ specifying the species for the sake of simplifying the notation.

\subsection{$\delta f$  and equations of motion discretization}
In \orb{}, the phase space is sampled using a set of $N$ markers that are distributed according to a
function $g(z, t)$ which is discretized as
\begin{equation}
  \label{eq:g_distrib}
  g(z, t) \simeq \sum_{i = 1}^N \frac{\delta\bl z - z_i(t)\br}{\jac_z},
\end{equation}
where $\delta[x]$ is the Dirac distribution, $z$ is a set of generalized phase-space coordinates,
$z_i(t)$ is the orbit of the i-th marker in phase space, and $\jac_z$ is the Jacobian associated
with the coordinates $z$ of the phase space. Even though the choice of the distribution function
$g(z, t)$ is not constrained, we make the convenient choice of using a distribution satisfying
\begin{equation}
  \label{eq:g_choice}
  \frac{\dd g}{\dd t}(z, t) = 0,
\end{equation}
where the $\dd/\dd t$ operator is the collisionless total time derivative defined by the general
Vlasov equation, \eqref{eq:Vlasov_general}. Both background and perturbed distribution functions
can be linked to the marker distribution by the weight fields $W(z, t)$ and $P(z, t)$:
\begin{align}
  \label{eq:marker}
  f_0 &= P(z, t)g(z, t)\simeq P(z, t)\sum_{i = 1}^N \frac{\delta\bl z - z_i(t)\br}{\jac_z} = \sum_{i = 1}^N
        P(z_i(t), t)\frac{\delta\bl z - z_i(t)\br}{\jac_z} = \sum_{i = 1}^N p_i(t)
        \frac{\delta\bl z - z_i(t)\br}{\jac_z}, \\
  \label{eq:df}
  \delta f &= W(z, t)g(z, t)\simeq W(z, t)\sum_{i = 1}^N \frac{\delta\bl z - z_i(t)\br}{\jac_z} = \sum_{i = 1}^N
  W(z_i(t), t)\frac{\delta\bl z - z_i(t)\br}{\jac_z} = \sum_{i = 1}^N \delta w_i(t) \frac{\delta\bl z - z_i(t)\br}{\jac_z},
\end{align}
where $p_i(t)=P(z_i(t), t)$ and $\delta w_i(t)=W(z_i(t), t)$ are the marker weights representing
respectively the amplitude of $f_0$ and $\delta f$ carried by each marker. The distribution
functions are normalized such that
\begin{equation}
  \int f(z, t) \jac(z) \dd z = N_{\text{ph}},
\end{equation}
where $N_{\text{ph}}$ is the physical number of particles in the system. Note that the coefficient
$N_{\text{ph}}/N$ is hereafter included in the weights such that $p_i(t) \equiv (N_{\text{ph}}/N) p_i(t)$ and  $\delta w_i(t) \equiv (N_{\text{ph}}/N) \delta w_i(t)$.

\subsubsection{Solving for the collisionless dynamics}
According to the time splitting approach, the collisionless dynamics is solved first using the
standard $\delta f$ or the direct $\delta f$ \cite{Allfrey2003} methods. For the standard $\delta f$
the time evolution of a marker $\delta w_i$ is given by
\begin{equation}
  \label{eq:w_evo}
  \frac{\dd}{\dd t} \delta w_i(t) =  \frac{\dd}{\dd t} W(z_i(t), t) =  \frac{\dd}{\dd t} \bl\frac{\delta f(z, t)}{g(z, t)}
  \br = \frac{1}{g(z, t)} \frac{\dd}{\dd t}\delta f(z, t) - \frac{\delta f(z, t)}{g(z, t)^2}\frac{\dd}{\dd t}g(z, t).
\end{equation}
The last term cancels out due to the choice of the distribution function $g(z, t)$,
\eqref{eq:g_choice}. The total distribution function $f$ being constant along collisionless
trajectories in phase space, the evolution equation of $\delta w_i$, \eqref{eq:w_evo}, can be written
as
\begin{equation}
  \label{eq:w_evo2}
  \frac{\dd}{\dd t} \delta w_i(t) = -g(z, t)\frac{\dd}{\dd t} f_0(z_i(t)) = - p_i(t) \frac{1}{f_0(z_i(t))}\frac{\dd}{\dd t} f_0(z_i(t)).
\end{equation}
Similarly, an equation for the $p_i$ weight can also be derived:
\begin{equation}
    \label{eq:p_evo2}
  \frac{\dd}{\dd t} p_i(t) = g(z, t)\frac{\dd}{\dd t} f_0(z_i(t)) = p_i(t) \frac{1}{f_0(z_i(t))}\frac{\dd}{\dd t} f_0(z_i(t)).
\end{equation}
In \orb{}, both equations are solved using a RK4 scheme and the particles are pushed using RK4
approximation of the particle's equations of motion.

On the other hand, the direct $\delta f$ method exploits the invariance of the total distribution
function $f$ along the nonlinear collisionless trajectories; this property is not ensured in the
linear and/or neoclassical limits. It allows one to directly evaluate the weights without
numerically solving a differential equation. Adding Eqs.~\neqref{eq:w_evo2} and \neqref{eq:p_evo2}
leads to
\begin{equation}
  \label{eq:f_const}
  \frac{\dd}{\dd t} (\delta w_i(t) + p_i(t)) = 0 \quad\Longrightarrow\quad \delta w_i(t) + p_i(t) = \delta w_i(t_0) +
  p_i(t_0), \forall t,
\end{equation}
which comes from the invariance of both $f$ and $g$ distribution functions. Furthermore, rewriting
\eqref{eq:p_evo2}, we find

\begin{equation}
  \label{eq:p_int}
  \frac{\dd}{\dd t}\bl\ln\pl \frac{p_i(t)}{f_0(z_i(t))}\pr\br = 0 \quad\Longrightarrow\quad \frac{p_i(t)}{f_0(z_i(t))} = \frac{p_i(t_0)}{f_0(z_i(t_0))}.
\end{equation}
The direct $\delta f$ algorithm consists of first evaluating the $p_i(t)$ weight using
\eqref{eq:p_int} and then computing the $\delta w_i(t)$ weight using \eqref{eq:f_const}. Note that
whatever the $\delta f$ method used, if the collisionless limit is considered only the $\delta w_i$
weights are required since the distribution $g(z, t)$ is invariant along the marker
trajectories. Indeed, inserting \eqref{eq:p_int} into \eqref{eq:w_evo2} gives
\begin{equation}
  \frac{\dd}{\dd t} \delta w_i(t)  = -  \frac{p_i(t_0)}{f_0(z_i(t_0))}\frac{\dd}{\dd t} f_0(z_i(t)).
\end{equation}
Therefore, we do not need to explicitly evolve $p_i(t)$.

\subsubsection{Solving for the collisional dynamics}
The collision operators are derived assuming linearization with respect to a local Maxwellian
distribution. However, \orb{} is typically operated using the canonical background Maxwellian
distribution in order to keep the background distribution in equilibrium in the collisionless
gyrokinetic equation. Upon entering the collisions module, the weights are converted to represent
the perturbation from a local Maxwellian background distribution, and are reverted when leaving
it. In this section, $f_0$ and $\delta f$ always refer to these converted distributions, \ie{}
$f_0 = f_\text{L}$ and $\delta f = f - f_\text{L}$. At each time step, the collision operators are
applied sequentially after the collisionless dynamics.

The electron-ion collision operator and the test-particle component of the intraspecies collision
operator are applied using a Langevin approach. In the gyrokinetic framework, this corresponds to
randomized ``kicks'' made in the velocity space.

For electrons colliding on ions, \eqref{eq:coll6} is reformulated in a spherical coordinate system
in velocity space with radius $r$, polar angle $\theta$, and azimuthal angle $\alpha$ in which the
incoming electron's velocity corresponds to $\theta = 0$. Then coming back in the \orb{} set of
coordinates, the outgoing trajectory of the electron is
\begin{align}
  v_{||,\textrm{out}} &= v_\textrm{in} \bl - \sin(\Delta \theta) \sin (\alpha_\textrm{out}) \sqrt{1 - \xi_{\textrm{in}}^2} + \xi_{\textrm{in}} \cos(\Delta \theta) \br, \\
  v_{\perp,\textrm{out}}^2 &= v_\textrm{in}^2 - v_{||,\textrm{out}}^2,
\end{align}
where $\Delta \theta = 2 R \sqrt{\nu_{\text{ei}}(v) \Delta t}$, where $R$ is a random sample of a
PDF with mean 0 and variance 1 and $\alpha_\textrm{out}$ is a random sample of a uniform
distribution between 0 and $2\pi$. Note that the energy is exactly conserved by this procedure as in
the original model.

Applying a similar approach for the ``test-particle'' self-collisions, \eqref{eq:coll1} yields the
following outgoing particle trajectory:
\begin{align}
v_{||,\textrm{out}} &= \frac{1}{v_{\textrm{in}} } \bl - \Delta v_y v_{\perp,\textrm{in}} + (v_{\textrm{in}} + \Delta v_z) v_{||,\textrm{in}}\br, \\
v_{\perp,\textrm{out}}^2 &= \Delta v_x^2 + \frac{1}{v_{\textrm{in}}^2} \bl\Delta v_y v_{||,\textrm{in}} + (v_{\textrm{in}} + \Delta v_z) v_{\perp,\textrm{in}} \br^2,
\end{align}
where $\Delta v_x$, $\Delta v_y$, and $\Delta v_z$ are the particle's change in velocity. The unit
vector $\hat{z}$ is in the direction of the incoming particle's velocity. These kicks are described by
\begin{align}
\Delta v_x &= v_{\text{th}} \sqrt{\frac{K(v) \bar{\nu} \Delta t}{2}} R_1,\\
\Delta v_y &= v_{\text{th}} \sqrt{\frac{K(v) \bar{\nu} \Delta t}{2}} R_2,\\
\Delta v_z &= -H(v) v \bar{\nu} \Delta t + v_{\text{th}} \sqrt{H(v) \bar{\nu} \Delta t} R_3,
\end{align}
where $R_1$, $R_2$, and $R_3$ are again independent random numbers sampled from a PDF with mean 0
and variance 1. The marker's parallel velocity $v_{||}$ and magnetic moment $\mu$ are then updated
accordingly.

It can be shown \cite{Vernay2010} that the evolution of the marker weight $\delta w_r$ due to collisions
can be expressed as
\begin{equation}
\frac{\dd}{\dd t} \delta w_r(t) = -p_r(t) \left.\frac{C[\delta f, f_{\text{L}}]}{f_{\text{L}}} \right\rvert _{[z_r(t),t]},
\end{equation}
where $z_r(t)$ is the marker position after the ``test-particle'' kicks. At this point, the
``background-reaction'' operator is slightly modified so as to ensure perfect conservation of mass,
momentum and energy inside each bin of space $\alpha$:
\begin{equation}
  \Delta \delta w_r(t) = -\frac{p_r}{n_\alpha} \bl\pl 1-3\sqrt{ \pi} G(x)\pr\Delta N_\alpha + 6 \sqrt{\pi}  H(x) \frac{\Delta P_{||,\alpha} v_{||\text{out},r}}{v_{\text{th},\alpha}^2}  + \sqrt{ \pi} G(x) \frac{\Delta E_\alpha}{v_{\text{th},\alpha}^2} \br,
\end{equation}
where $\Delta N_\alpha$, $\Delta P_{||\alpha}$ and $\Delta E_\alpha$ corresponds respectively to the
change in mass, momentum and energy in the bin $\alpha$ caused by the ``test-particle''
operator. This procedure ensures the conservation of mass, momentum and energy of the $\delta f$ to
machine precision.

\subsubsection{Particle loading}
At the beginning of a simulation, the markers are loaded in phase space using a Halton-Hammersley
sequence \cite{Halton1960,Hammersley1960} and according to the distribution function
$g(z, t=0) = f_s(s)f_v(v_\|, v_\perp)$, where $f_s(s)$ and $f_v(v_\|, v_\perp)$ define respectively
the radial and velocity sampling distributions. In \orb{}, the spatial sampling is defined by the
\textit{specified loading} distribution function
$f_s(s) = 1-f_\text{g}+f_\text{g}\exp\bl\pl s- s_0\pr^2/\Delta s^2\br$, where $f_\text{g}\in[0, 1]$,
$s_0$, and $\Delta s$ are input parameters. In velocity space $(v_\|, v_\perp)$, the markers are
uniformly distributed in $|v|^2$ or $|v|^3$ with a cut-off at $|v| = \kappa_vv_{\text{th},s}$, where
$\kappa_v$ is an input parameter usually set at $\kappa_v=5$.

For the marker weight initialization, two main schemes are implemented. The first option is a
\textit{white noise initialization} defined by
\begin{equation}
  \delta w_i(t_0) = A (2Q_i-1) p_i(t_0),
\end{equation}
where $Q_i$ is a quasi-random number in $[0, 1]$ given by the i-th term of a van der Corput sequence
\cite{Corput1935} and $A$ the maximum amplitude given as an input parameter, typically of the order
of $A \sim 10^{-3}-10^{-5}$. The disadvantage of this scheme is that the initial density or current
perturbation is inversely proportional to the number of particles and the time until physical modes
emerge from the initial state is roughly proportional to the number of particles. To accelerate the
mode development, the \textit{mode initialization} can be used. It consists in initializing a number
of Fourier modes:
\begin{equation}
  \delta w_i(t_0) = \frac{A_0 p_i(t_0)}{(m_2-m_1+1)(n_2-n_1+1)}\left|\frac{T(s_0)}{\grad T(s_0)} \right|
  \times \left|\frac{T(s_i(t_0))}{\grad T(s_i(t_0))} \right|
  \sum_{m=m_1}^{m_2}\sum_{n=n_1}^{n_2}\cos(m\ts_i(t_0)-n\varphi_i(t_0)),
\end{equation}
where $A_0$, $n_1$, $n_2$, $m_1$, $m_2$ are input parameters. Typically, for linear simulations of
microinstabilities with a toroidal mode number~$n_0$, it is convenient to use $n_1 = n_2 = n_0$
and $m_1 = m_2 = -n_0q(s_0)$ as modes are almost aligned with the magnetic field
lines. Finally, whatever initialization is used, the initial average value of the weights is set to
zero:
\begin{equation}
  \frac{1}{N}\sum_i^{N}\delta w_i(t_0) = 0.
\end{equation}

As mentioned in section \ref{sec:definitions} the markers are pushed in toroidal magnetic
coordinates $(s, \ts, \varphi)$. To avoid the singularity that would appear in the equations of
motion at the magnetic axis, the coordinate system is changed to
$(\xi, \eta, \varphi) = (s\cos\ts, s\sin\ts, \varphi)$ near the axis. All equilibrium quantities for
both \textit{ad-hoc} and MHD equilibria are loaded on an $(R, Z)$ grid and are linearly interpolated
to an $(s, \ts)$ grid. Markers that exit the radial domain at $s>1$ are reflected back into the
plasma at a position which conserves toroidal momentum, the particle energy, and the magnetic moment
but with a null weight to avoid unphysical accumulation of perturbed density at the radial edge.

\subsection{Quasineutrality and Ampère equations}
\label{sec:QNE}
In \orb{} the quasineutrality and Ampère equations are solved using the Galerkin method and linear,
quadratic, or cubic B-splines finite elements defined on a $(N_s, N_{\ts}, N_\varphi)$ grid. The
perturbed fields $\phi$ and $A_\|$ hereafter noted $\Psi = \{\phi,A_\|\}$ are discretized as
follows:
\begin{equation}
  \label{eq:discretization}
  \Psi(\vec{X}, t) = \sum_\mu \Psi_\mu(t)\Lambda_\mu(\vec{X}),
\end{equation}
where $\{\Psi_\mu(t)\}$ are the field coefficients and $\{\Lambda_\mu(\vec{X})\}$ are a tensor
product of 1D B-splines of degree $p=\{1, 2, 3\}$, $\Lambda_\mu(\vec{X}) =
\Lambda_j^p(s)\Lambda_k^p(\ts)\Lambda_l^p(\varphi)$, with $\mu = (j, k, l)$.

Using the decomposition defined in Eqs.~\neqref{eq:df} and \neqref{eq:discretization}, and setting
the test functions $\hat{\phi}_1 = \Lambda_\nu(\vec{X})$, $\nu = (j', k', l')$ of the variational
forms of the quasineutrality and Ampère equations, Eqs.~\neqref{quasin_full} and
\neqref{GK_Ampere_full}, leads to a linear system of the form
\begin{equation}
  \label{eq:linsys}
  \sum_\mu A_{\mu\nu}\Psi_\mu(t) = b_\nu(t),
\end{equation}
where $A_{\mu\nu}$ and $b_\nu$ are respectively a real symmetric positive-definite matrix and a
vector that are defined by the physical models used in the quasineutrality and Ampère equations. Due
to the finite support of the B-splines, the matrix $A_{\mu\nu}$ is usually a block matrix composed
of banded submatrices. For the sake of illustration, we show here the linear system for the case of a
single species plasma in the limit of adiabatic electrons with the long wavelength approximation for
the ion polarization density:
\begin{align}
  \label{eq:Amunu}
  A_{\mu\nu}^{\text{LWA,adiab}} &= \int \bl\frac{e n_0(\psi)}{T_\text{e}(\psi)}\pl
  \Lambda_\mu(\vec{X})\Lambda_\nu(\vec{X})-\bar{\Lambda}_\mu(s)\bar{\Lambda}_\nu(s)\pr +
                           \frac{n_0(\psi)m_\text{i}}{B^2}\gradperp\Lambda_\mu(\vec{X})\cdot\gradperp\Lambda_\nu(\vec{X})\br
                                  \dd V,\\
  \label{eq:bnu}
  b_\nu(t) &= \sum_{p=1}^N \frac{\delta w_p(t)}{2\pi}\int_0^{2\pi}\dd\alpha\Lambda_\nu\pl\vec{X}_p+\vec{\rho}_{\text{L},p}(\alpha)\pr,
\end{align}
where $\vec{\rho}_{\text{L},p}$ is the Larmor radius of a particle $p$. Here, the perpendicular
gradient is approximated by the poloidal gradient, \ie{}
$\gradperp \simeq \grad_{\text{pol}} = \grad s \pard{}{s}+\grad\ts\pard{}{\ts}$. Note that the
expression for $b_\nu(t)$, \eqref{eq:bnu}, is independent of the choice of coordinates. This is due
to the particle representation of $\delta f$, \eqref{eq:df}, and the Galerkin finite element method
based on the variational form of the field equations, Eqs~\neqref{quasin_full} and
\neqref{GK_Ampere_full}. This is very convenient practically as the charge deposition is totally
transparent from the choice of the coordinates system, which greatly simplifies the numerical
implementation. A more complete description of the discretized Poisson equation for arbitrary
wavelengths can be found in \cite{Dominski2017}.

In \orb{}, the linear system of equations, \eqref{eq:linsys}, is solved in discrete Fourier space
\cite{McMillan2010} using the {\sc Fftw} library \cite{Frigo2005} and a direct solver from the {\sc
  Lapack} library \cite{Anderson1999}. The Fourier representation of the fields in an axisymmetric
magnetic confinement device is convenient because of the double periodicity in the toroidal and
poloidal directions of the flux surfaces. Furthermore, the modes of interest, \eg{} drift-wave type
and Alfvén waves, are typically almost aligned with the magnetic field lines and can be described
with just a small set of Fourier coefficients, which greatly decreases the numerical cost as
compared to solving the system in direct space. Noting $\mathcal{F}$ the double discrete Fourier
transform on both poloidal and toroidal directions, the linear system of equations
\neqref{eq:linsys} becomes
\begin{equation}
  \label{eq:Fouriersys}
  \sum_\mu\mathcal{F}A_{\mu\nu}\mathcal{F}^{-1}\mathcal{F}\Psi_\mu = \mathcal{F}b_\nu,
\end{equation}
\begin{equation}
  \mathcal{F}A_{\mu\nu}\mathcal{F}^{-1} = \hat{\hat{A}}_{(j,j')}^{(n,m),(n',m')},
\end{equation}
\begin{equation}
  \mathcal{F}\Psi_\mu = \hat{\hat{\Psi}}_j^{n,m},
\end{equation}
\begin{equation}
  \mathcal{F}b_\nu = \hat{\hat{b}}_{j'}^{n,m'},
\end{equation}
where $n$ and $m$ are respectively the toroidal and poloidal Fourier mode numbers.

Due to the axisymmetry of the system, the toroidal direction can be decoupled from the
others with $n = n'$ \cite{Tran1999}:
\begin{equation}
  \sum_j\sum_m \hat{\hat{A}}_{(j,j')}^{(n,m),(n,m')}\hat{\hat{\Psi}}_j^{n,m} =
  \frac{\hat{\hat{b}}_{j'}^{n,m'}}{M^{n, p}} \quad \forall n,
\end{equation}
where the matrix $M^{n, p}$ is defined by
\begin{equation}
  M^{n, p}=\sum_{l'=1}^{N_\varphi}\int\dd\varphi\Lambda_{l'}^p(\varphi)\Lambda_{l}^p(\varphi)\exp\bl\frac{2\pi
  {\rm i}}{N_\varphi}(l'-l)\br,
\end{equation}
and can be computed analytically for any B-spline of order $p$.

The matrix $A_{\mu\nu}$ and the right-hand side $b_\nu$ are modified such that the following
boundary conditions are used. At the magnetic axis the \textit{unicity condition} is applied,
$\Psi(s=0, \ts, \varphi, t) = \Psi(s=0, \ts=0, \varphi, t), \forall \ts$. At the outer radial edge,
Dirichlet boundary conditions are applied, $\Psi(s=1, \ts, \varphi, t) = 0$. Note that \orb{} can
also be run in an annulus, \ie{} $s\in[s_{\text{min}}, s_{\text{max}}]$, with $s_{\text{min}}>0$ and
$s_{\text{max}}<1$, for which case Dirichlet boundary conditions are applied on both edges. For the
quasineutrality equation with polarization density at all orders, the equation is integral and no
Dirichlet boundary conditions are applied \cite{Dominski2017}.

\subsubsection{Gyroaveraging}
For all gyroaveraging operations, the plane of the Larmor ring is approximated to lie in the
poloidal plane. The number of gyropoints can be either fixed or determined by an adaptive scheme:
a fixed number of Larmor points is used for all the particles having a Larmor radius smaller or
equal to the thermal Larmor radius and the number of points increases linearly for larger Larmor
radii. Usually, a fixed number of 4 gyropoints is sufficient for perturbations up to
$k_\perp\rho_{\text{L}}\sim 1$. However, using the adaptive scheme reduces the noise as it acts as a
Bessel filter smoothing out shorter wavelength fluctuations \cite{Hatzky2002}.

In magnetic coordinates the positions of the gyropoints are parametrized using the gyroangle
$\alpha$:
\begin{equation}
  \label{eq:gyroaverage}
  \vec{x}(\alpha) = \vec{X} + \vec{\rho}(\alpha) = \vec{X} + \rho \frac{\grad s}{|\grad
    s|}\cos\alpha + \rho \frac{\vec{b}\times\grad s}{|\vec{b}\times\grad s|} \sin\alpha,
\end{equation}
where $\vec{X}$ is the position of the guiding center.

The gradients of gyroaveraged electric potential, $\grad\left\langle\phi_1\right\rangle$, is
defined as
\begin{equation}
%  \label{eq:gyroaverage}
  \grad_{\vec{X}}\left\langle\phi_1\right\rangle =
  \frac{1}{2\pi}\oint_0^{2\pi}\grad_{\vec{X}}\phi_1\pl\vec{X}+\vec{\rho}\pr \dd\alpha,
\end{equation}
where the subscript $\vec{X}$ stands for the gradient with respect to the gyrocenter coordinates and
$\alpha$ is the gyroangle. We define a new set of coordinates
$\bar{\vec{X}} = \pl\bar{R}, \bar{Z}\pr = \pl R + \rho\cos\alpha, Z + \rho\sin\alpha\pr=\vec{X}+\vec{\rho}$
representing the position of the particle on the gyro-ring in the poloidal plane where $\bar{R}$ is
in the direction of the major axis and $\bar{Z}$ is in the direction of the vertical axis. Using the
chain rule, the term $\grad_{\vec{X}}\phi_1$ from \eqref{eq:gyroaverage} can be written as
\begin{equation}
  \label{eq:gyroaverage2}
  \grad_{\vec{X}}\phi_1\pl\vec{X}+\vec{\rho}\pr = \grad_{\bar{R}}\phi_1-\frac{\rho}{2}\pl
  \frac{\partial\phi_1}{\partial\bar{R}}\cos\alpha+\frac{\partial\phi_1}{\partial
      \bar{Z}}\sin\alpha\pr \frac{\grad_{\vec{X}}B}{B}.
\end{equation}
A similar procedure is done for $\grad\left\langle A_{1\|}\right\rangle$. In \orb{},
\eqref{eq:gyroaverage2} can be either directly evaluated as in \cite{Hatzky2019} or approximated by
neglecting the second term, leading to $\grad_{\vec{X}}\phi_1 \approx \grad_{\bar{R}}\phi_1$.

\subsubsection{Fourier filter}
Typical modes of interest, \eg{} drift waves and low-frequency Alfvén waves, are mainly aligned with
the magnetic field lines, \ie{} they have $m\approx nq(s)$. Due to this strong anisotropy, only a
small set of $(n, m)$ Fourier coefficients is required to describe the modes as their amplitude
rapidly decreases away from $m=nq(s)$ \cite{Jolliet2007}. It is then beneficial to filter out all
the non physically relevant Fourier modes in order to reduce the sampling noise and maximize the
timestep size. The filter is applied on the Fourier coefficients of the perturbed density and
current to filter out all the non physical modes introduced by the charge and current depositions:
\begin{equation}
  \tilde{b}_{(j,k,l)}= \sum_{n,m}f_{j,n,m}\hat{\hat{b}}_{j}^{n,m}e^{im\ts_k}e^{in\varphi_l},
\end{equation}
where $f_{j,n,m}$ is the Fourier filter that in general depends on the radius, and the poloidal and
toroidal mode numbers.

Two different filters are used successively. First, a \textit{rectangular filter}, which is the most
simple one, is applied such that all the modes outside of the window
$[\nmin, \nmax]\times[\mmin, \mmax]$ specified in input are filtered out. This filter is not
sufficient as it keeps modes with $k_\|/k_\perp$ much bigger than $\rho^\star$, which is inconsistent
with the gyrokinetic ordering \cite{Jolliet2007}. Since the modes of interest are mainly aligned
with the magnetic field, \ie{} they satisfy
$k_\|\rho_\text{i} = \bl m+nq(s)\br\bl q(s)r\br^{-1}\rho_\text{i}= \mathcal{O}\pl\rho^\star\pr$, a
second surface-dependent \textit{field-aligned filter} is applied. It consists in retaining only $m$
modes close to $-nq(s)$, \ie{} $m \in [nq(s)-\Delta m, nq(s)+\Delta m]$, where $\Delta m$ is an
input parameter specifying the width of the filter. With this field-aligned filter, the maximum
value of $|k_\||$ represented is $|k_\||_{\text{max}} = |\Delta m|/qR$. Since
$|k_\||_{\text{max}}\rho_L$ scales with $\rho^\star$, the value of $\Delta m$ required to describe
all physically relevant modes is invariant with the system size. Typically, a value of $\Delta m =5$
is sufficient \cite{McMillan2010}. In summary, for each mode $n\in[\nmin, \nmax]$ only the modes
$m\in[\mmin, \mmax]\cap[-nq(s)\pm\Delta m]$ are retained.

\subsection{Noise control techniques}
Due to the finite number of markers used to sample the phase space, PIC simulations are subject to
noise accumulation deteriorating the signal quality and forbidding long simulations without noise
control techniques. All the difficulty of such noise-reducing schemes is to actually control the
weight growth without creating severe non-physical artifacts. In this section we present the
different noise control schemes implemented in \orb{}.

\subsubsection{Krook operator}
\label{sec:Krook}
The Krook operator implemented in \orb{} \cite{McMillan2008} is a source term which weakly
damps the non axisymmetric fluctuations without significantly affecting the zonal flows. This is
done via a correction term that also allows one to conserve various moments by projecting out some
components of the source. The Krook noise-control term, $S_{\text{K}}^{\text{NC}}$, is composed of a relaxation
term and its correction $S_\text{K}^{\text{corr}}$:
\begin{equation}
  \label{eq:KrookTotal}
  S_\text{K}^{\text{NC}} = -\gamma_\text{K} \delta f + S_\text{K}^{\text{corr}},
\end{equation}
\begin{equation}
  \label{eq:KrookCorr}
  S_\text{K}^{\text{corr}} = \sum_{i=1}^{N_{\text{mom}}}g_i(s)M_if_0,
\end{equation}
where $\gamma_\text{K}$ is the Krook damping rate. The correction term is a sum over the
$N_{\text{mom}}$ moments $M_i$ one wishes to conserve on a flux-surface average. Typically, in
\orb{}, the moments that can be conserved are the density, parallel velocity, zonal flows, and
kinetic energy. They are respectively defined by
$M_i = \{1, v_\|, v_\|/B-\widetilde{(v_\|/B)}, \mathcal{E}_\text{K}\}$, where the tilde represents
the bounce average and $\mathcal{E}_\text{K}$ is the kinetic energy of a particle. The coefficients
$g_i(s)$ are defined such that there is no contribution of the source to a given moment $M_j$, \ie{}
\begin{equation}
  \label{eq:corr}
  \overline{\int \dd W M_j S_\text{K}^{\text{NC}}} = 0,
\end{equation}
where the over bar represents the flux-surface average. Injecting the definition of the Krook source
term, \eqref{eq:KrookTotal}, in \eqref{eq:corr} leads to a linear system of equations that
is solved at each time step to find the coefficients $g_i(s)$:
\begin{equation}
  \sum_i^{N_{\text{mom}}}S_{ij}(s, t)g_i(s, t) = \delta S_j(s, t),
\end{equation}
with
\begin{equation}
  S_{ij}(s, t) = \overline{\int \dd W M_j(\vec{X}, v_\|, \mu)M_i(\vec{X}, v_\|, \mu)f_0(\vec{X}, v_\|, \mu)},
\end{equation}
\begin{equation}
  \delta S_{j}(s, t) = \gamma_K\overline{\int \dd W \delta f(\vec{X}, v_\|, \mu, t)M_j(\vec{X}, v_\|, \mu)}.
\end{equation}
Note that the flux-surface average is numerically represented by a binning of the markers in the
radial direction. This implies that the conservation in ensured only on average across each radial
bin.

As already mentioned, the noise control should not affect significantly the turbulence. To this end,
values of the order of one tenth of the maximum linear growth rate are usually used for the Krook
damping rate. In this way, the linear phase is not substantially modified and a high signal-to-noise
ratio can be obtained. On the other hand, this noise control technique cannot be used with
collisions when the damping rate $\gamma_\text{K}$ is comparable to the collision frequency thus masking
the effect of collisions.

By construction, the Krook operator damps the fluctuations to restore the full distribution function
to its initial state. If the kinetic energy is not conserved while conserving the other moments, it
allows one to run temperature gradient-driven simulations by acting as an auto-regulated heat
source while allowing for density and flow profile unconstrained evolution.

\subsubsection{Coarse-graining}

Coarse-graining \cite{Chen2007} is an additional noise-control method implemented in \orb{}
\cite{Vernay2012} to reduce the problems of weight-spreading and filamentation of the distribution
function, that lead to large mean squared particle weights. The idea is essentially to dissipate
fine-scale structures of the distribution function in phase space, as represented by the marker
weights. This is an improvement in comparison to the Krook operator, which only preserves certain
moments of the zonal distribution function but otherwise somewhat indiscriminately damps the whole
distribution function; the Krook operator, for example, was found to be unsuitable for neoclassical
studies.

In an Eulerian code, phase space dissipation is often implemented as a hyper-viscosity on the grid in the
five spatial and velocity directions. As the PIC approach does not involve a phase-space grid, we need
an alternative method to smooth the weights of nearby markers.

Computationally, the method consists of binning the particles in field-aligned grid cells in phase
space, and then reducing the deviation of particle weights in the grid cell from their average
value. To avoid smoothing structures at the turbulence scale too strongly, the bins must be small
compared to typical length and velocity scales; on the other hand the bins need to frequently
contain more than one marker for this procedure to be effective. Field-aligned bins are used because
the distribution function varies much more rapidly perpendicular to the field line than across it.

The bins are volumes in a block-structured Cartesian mesh in coordinates
$(s,z,\ts,\lambda,\epsilon)$, with the number of bins uniform in each direction, except that the
number of bins in the $\ts$ direction is proportional to $s$, so that the spatial volume of bins is
roughly constant. The coordinate $\epsilon$ is the particle kinetic
energy, $\lambda$ is the pitch angle, and $z$ is a field-line label that is computed as
\begin{equation}
  z = \varphi - q(s) [\ts - \ts_0(\ts)],
\end{equation}
with $\ts_0$ the center of the bin in the $\ts$ direction. With this choice of $z$ we have a
field-aligned bin, but we also have $ z \sim \varphi $ if there are many bins in the $\ts$ direction
because $\ts_0(\ts)$ tends to $\ts$ for an infinite number of bins. This is useful because the
domain decomposition---discussed in details in Section~\ref{sec:parallelization}---means that
markers on a single processor have a small range of values of $\varphi$. Thus, the first step in the
binning computation is to distribute the markers according to $z$ and move them to this alternative
domain decomposition. In the $z$ decomposition, coarse-graining is local to each domain, so we do
not need to communicate quantities on the 5D coarse-graining mesh.

The number of bins in the $s$ and $z$ directions are the field mesh quantities $N_s$ and $N_\varphi$
respectively and the number of $\ts$, energy and pitch-angle bins are specified as input
parameters. To avoid excessive damping of zonal flows, around $32$ bins are needed in each of the
energy and pitch-angle directions. Often $16$ bins in the $\ts$ direction are sufficient to avoid
excessive damping of parallel structures.

The smoothing operation changes the particle weight $w$ by an amount
$ \mathcal{N} \Delta t \gamma_\text{cg} ( \bar{w} - w )$, where $\bar{w}$ is the average particle weight in
the bin, $ \mathcal{N}$ is the number of timesteps (of length $\Delta t$) between coarse-graining
operations, and $\gamma_\text{cg}$ is a parameter controlling the coarse-graining rate. In the large-marker
limit, this leads to a damping of fine-scale structures in the distribution function with a rate
$\gamma_\text{cg}$. Note, however, that in practice, typical runs have $0.1$ markers per bin, so that the
effective coarse-graining rate is lower than $\gamma_\text{cg}$ by a factor of $10$.

\subsubsection{Quad-tree particle-weight smoothing}

The grid-based coarse-graining procedure has the possible drawback of being inaccurate if the grid
of phase-space bins is too fine so that the local statistics is not good enough, or being very
diffusive if the grid is too coarse. An alternative procedure, gridless in velocity space and more
probabilistic in nature, has been proposed in \cite{Sonnendrucker2015} and implemented in \orb{}. It
consists in pairing neighbouring markers and replacing their weights by an average, weighted by a
function of their distance in velocity space. The way of computing the distance and the weight has
an influence on the diffusivity of the method, for this reason we use a procedure for pairing only
particles which are close enough. Since the gyrokinetic velocity space is 2D, the pairing procedure
is done using a quad tree algorithm: first, the particles are binned in the configuration space and
then, a quad tree procedure is applied to define regions in velocity space within which particles
will be paired. This works by subdividing recursively the 2D velocity space in four sub-boxes until
the number of particles in a sub-box is smaller than a given value set as an input parameter. At
this point, the particles within a sub-box are randomly paired and their weight is changed according
to the following procedure: for a pair of two markers with weights $\delta w_1$ and $\delta w_2$ and velocities
$\vec{v}_1$ and $\vec{v}_2$, the new weights are given by
\begin{align}
  w^{\rm new}_{\rm 1} &= \pl 1-\Gamma\pr w^{\rm
                          old}_{\rm 1} + \Gamma\,\bar{w}, \\
  w^{\rm new}_{\rm 2} &= \pl 1-\Gamma\pr w^{\rm
                          old}_{\rm 2} + \Gamma\,\bar{w},
\end{align}
with
\begin{align}
  \Gamma &= {\rm e}^{-\frac{\pl v_{1}^x-v_{2}^x\pr^2+\pl v_{1}^y-v_{2}^y\pr^2}{h_v^2}},\\
  \bar{w} &= \frac{w^{\rm old}_{\rm 1}+w^{\rm old}_{\rm 2}}{2},
\end{align}
where the $x$ and $y$ superscripts are used to identify the two dimensions of the velocity and the
$h_v$ parameter defines how strong is the smoothing procedure with respect to the distance
separating the pair of markers in velocity space. Note that, by construction, the smoothing
operation conserves the total weight, ensuring density conservation. By picking different pairs of
particles within the same quad tree sub-box, the smoothing operation can be applied several times
per timestep. Typically, one smoothing step is done at every timestep.

\subsubsection{Enhanced control variate}

The \orb{} code solves the uncoupled electromagnetic gyrokinetic equations in the
$p_{z}$-formulation, Eqs.~\neqref{quasin_full} and \neqref{GK_Ampere_full} and therefore includes
the cancellation problem \cite{Chen2001} which, if untreated, in practice limits the electromagnetic
simulations to very-low-beta cases, $\beta <\sqrt{m_{\rm e}/m_{\rm i}}$, where $\beta$ is the stored
kinetic energy divided by the magnetic field energy. Different methods mitigating this problem have
been developed for the particle-in-cell framework in
Refs.~\cite{Chen2003,Mishchenko2004a,Mishchenko2004,Hatzky2007,Hatzky2019} and for the Eulerian
approach in Ref.~\cite{Candy2003a}. In \orb{} the cancellation problem is treated
\cite{Mishchenko2017} using the enhanced control variate scheme presented in
\cite{Hatzky2007,Hatzky2019}. A further development of the mitigation schemes is given in
Refs.~\cite{Mishchenko2014a,Kleiber2016}, the so-called pullback mitigation based on the
mixed-variable formulation \cite{Mishchenko2014} of the gyrokinetic theory, has also been
implemented in \orb{} \cite{Mishchenko2018a}.
%Another mitigation technique has been considered in Ref.~\cite{Bao_Lin}.
Mitigation of the cancellation problem made possible the \orb{} electromagnetic simulations described in
Refs.~\cite{Biancalani2016,Biancalani2017}.

The enhanced control variate approach is based on the decomposition of the distribution function
into the so-called adiabatic and nonadiabatic parts introduced in \cite{Catto1981} while
constructing a perturbative procedure for the solution of the gyrokinetic Vlasov equation. The same
decomposition can be extracted via the pull-back transformation between the particle distribution
function and the reduced gyrokinetic distribution \cite{Brizard2007}. This transformation requires
that the equilibrium distribution commutes with the background dynamics,
\ie{} $\left\{f_{\text{eq}},H_0\right\}=0$. Furthermore, in \orb{} the distribution function is assumed to
be a canonical Maxwellian, \ie{} satisfying
\begin{equation}
  \frac{\dd f_{\text{eq},s}}{\dd H_0}=\frac{f_{\text{eq},s}}{T_s},
\end{equation}
where the temperature is defined as
\begin{equation}
T_s=\frac{1}{n_{0}}\int\dd W\ \frac{p_z^2}{2 m_s} f_{\text{eq},s}.
\end{equation}
In the enhanced-control-variate scheme, the perturbed distribution function is split according to
\begin{equation}
  \label{eq:distsplit}
f_{s,1}= G_s -\frac{f_{\text{eq},s}}{T_s}\left\langle H_1\right\rangle,
\end{equation}
where the first and second terms are respectively the nonadiabatic and adiabatic parts.

The cancellation problem is related to the coexistence of very large and very small quantities in
the variational form of the Amp\`ere equation \neqref{GK_Ampere_full}. To illustrate the problem,
let us consider a case with only one ion species and rewrite \eqref{GK_Ampere_full}. First, the
second and third integrals of \eqref{GK_Ampere_full} are the projections $\mathcal{J}_{s,\|}$of the
ion and electron currents onto the basis function $\widehat{A}_{1\|}$:
\begin{align}
  \label{eq:currentsi}
  \left\langle \mathcal{J}_{\text{i},\|}\right\rangle & \equiv \int \, \dd V\  \left\langle j_{\text{i},\|}\right\rangle \widehat{A}_{1\|}= \int \ \dd \Omega\ f_s \
                                              \frac{q_sp_{z}}{m_s}\left\langle\widehat{A}_{1\|}\right\rangle,
  \\
    \label{eq:currentse}
  \mathcal{J}_{\text{e},\|} &\equiv \int \, \dd V\ j_{\text{e},\|}\widehat{A}_{1\|} = \int \ \dd \Omega\ f_\text{e} \ \frac{e\,p_{z}}{m_\text{e}}\ \widehat{A}_{1\|}.
\end{align}
Then, the fourth integral and the first term of the fifth integral of \eqref{GK_Ampere_full} are the
so-called skin terms and can be written as
\begin{equation}
  \label{eq:skin_terms}
  \int \dd \Omega\ f_{\text{eq,}s}\ \pl\frac{4\pi q_s^2}{m_s c^2}A_{1\|}\widehat{A}_{1\|}\pr = \frac{\beta_s}{\rho_{\text{th,}s}^2}\int
    \dd V \, A_{1\|}\widehat{A}_{1\|},
\end{equation}
where one defines $\beta_s = 4\pi n_s T_s/B^2$ and $\rho_{\text{th,}s}$ is the thermal Larmor
radius. Finally, the remaining terms of \eqref{GK_Ampere_full} are combined to form
\begin{align}
  &\int \dd V\ \bm\nabla_{\perp}
  A_{1\|}\cdot\bm\nabla_{\perp}\widehat{A}_{1\|}
  + \int \dd \Omega\ f_{\text{eq,i}}\
  \frac{2\pi \mu}{B}
  \left[
    A_{1\|}\bm\nabla_{\perp}^2\widehat{A}_{1\|}+
    \widehat{A}_{1\|}\bm\nabla_{\perp}^2A_{1\|}
    \right]\\
  \label{eq:beta_term}
  =&  \int \dd V \, \bm{\nabla}_{\perp}\cdot\left[\left(1-\beta_\text{i}\right)\bm{\nabla}_{\perp}A_{1\|}\
    \widehat{A}_{1 \|}\right],
\end{align}
where the integration by parts has been used and the terms containing second order gradients of the
background quantities neglected. Putting Eqs.~\neqref{eq:currentsi}--\neqref{eq:beta_term} back into
\eqref{GK_Ampere_full} leads to
\begin{equation}
\frac{\beta_\text{i}}{\rho_\text{th,i}^2}\ \int \dd \Omega \ f_{\text{eq,i}}\
A_{1\|}\widehat{A}_{1\|}+\frac{\beta_\text{e}}{\rho_\text{th,e}^2}\ \int \dd \Omega\ f_{\text{eq,e}}\ A_{1\|}\widehat{A}_{1\|}-\int \dd \Omega\ \bm{\nabla}_{\perp}\bl\left(1-\beta_\text{i}\right)\bm{\nabla}_{\perp}A_{1\|}\ \widehat{A}_{1 \|}\br=\frac{4\pi}{c}\left(\left\langle \mathcal{J}_{\text{i}, \|}\right\rangle-\mathcal{J}_{\text{e},\|}\right).
\label{control_variate_amp}
\end{equation}
The two skin terms can become very large, especially for electrons, due to their small mass. They
cancel up to the second order FLR corrections with the adiabatic part of the currents
$\left\langle j_{\text{i},\|}\right\rangle$ and $j_{\text{e},\|}$. This can be seen by splitting the
currents into an adiabatic and nonadiabatic part using the splitting defined in \eqref{eq:distsplit}
and injecting them back into \eqref{control_variate_amp}.

The cancellation problem occurs in PIC simulations due to the different discretization of the
particles and fields: the currents are typically computed using the particles while the skin terms
are computed using the finite element grid. The terms to be cancelled are much larger in magnitude
than the remaining terms which are supposed to represent the physics. Therefore, the cancellation
must be numerically extremely accurate, otherwise the relevant signal is dominated by numerical
noise.

In \orb{}, the cancellation problem is mitigated discretizing the skin terms and the adiabatic part
of the currents in \eqref{control_variate_amp} with the same markers. The polarisation-current term,
$-\int \dd \Omega\ \bm{\nabla}_{\perp}\cdot\bl\left(1-\beta_\text{i}\right)\bm{\nabla}_{\perp}A_{1\|}\
  \widehat{A}_{1 \|}\br$, is discretized on the grid since it does not contribute to the
cancellation. This approach to the discretization is used in \orb{} in Ampère's law.

Ampère's law, \eqref{control_variate_amp}, is used to compute the parallel magnetic potential
$A_{1\|}$. Note that the non-adiabatic perturbed distribution function $G_s$ depends on $A_{1\|}$
which is unknown at this point of the computation. The solution is to use an easy-to-compute
estimator, $\hat{s}$, and solve iteratively for $A_{1\|}$. In \orb{}, the skin term
$(\beta_{\rm s}/\rho_{\rm s}^2) A_{1\|}$ is used as a simple estimator for the $A_{1\|}$-dependent
part of the distribution function. One reformulates Ampère's law using the estimator
$\hat{s}$:

\begin{equation}
  \label{eq:estimator}
  (\hat{s} + L) \, a = (j - s a) + \hat{s} a,
\end{equation}
where $a$ is the discretized magnetic vector potential component, $s$ and $L$ are respectively the
discretized skin terms and Laplacian term, and $j$ represents the sum over the species of the
discretized currents. For a good estimator, a small parameter $\| \hat{s} - s \| = {\cal O(\veps)}$
can be introduced to expand the vector potential,
$a = a_0 + \veps a_1 + \veps^2 a_2 + \mathcal{O}(\veps^3)$. Ampère's law is then solved iteratively
order by order in $\veps$:
\begin{align*}
  (\hat{s} + L) \, a_0 &= j, \\
  (\hat{s} + L) \, a_1 &= ( \hat{s} - s) \, a_0, \\
  &\ldots
\end{align*}
In practice, for typical production runs, less than 10 iterations are necessary. In \orb{}, the
estimator is expressed using the finite elements
$\hat{s}_{kl} = \int \beta_s/\rho_s^2 \; \Lambda_k(\vc{x}) \Lambda_l(\vc{x}) \, \df^3 x$.  The
marker-dependent part of the right-hand side of the iterative scheme is written as the enhanced
control variate: %\cite{Hatzky_2007}
%\begin{equation}
%j_{k} - s_{kl} a_{l}^{n-1} = \sum_{\nu=1}^{N_{\rm p}} p_{z\nu} \, \left( w_{\nu} +
%\frac{q_s \gav{A_{1\|}^{(n-1)}}}{m_s} \, \pard{f_{eq}}{p_{z}}(Z_{\nu}) \; \zeta_{\nu} \right) \gav{\Lambda_k}_{\nu}
%\end{equation}
\begin{equation}
j_{k} - s_{kl} a_{l}^{n-1} = \sum_{\nu=1}^{N_{\rm p}} p_{z\nu} \, \left( \delta w_{\nu} +
\frac{q_s p_z\ \gav{A_{1\|}^{(n-1)}}}{m_s} \, \frac{f_{\text{eq},s}}{T_s}(Z_{\nu}) \; \zeta_{\nu} \right) \gav{\Lambda_k}_{\nu}.
\end{equation}
The same enhanced control variate is used also for the perturbed particle density. In practice, it
results in a straightforward and computationally cheap modification of the charge and current
assignment routines in \orb{}.

\subsection{Heating operators}
A primary goal of simulating the full plasma core (by contrast to a local approach) is to examine
the self-consistent evolution of plasma profiles in the presence of both turbulence-driven
transport, and external sources, which are each generally of equal importance.  In practice, even
for running global simulations where realistic global profile evolution is not of interest, it is
generally inconvenient to run simulations without a heat source: if the goal is to look at transport
properties at a specific temperature gradient, simulations where the temperature gradient relaxes
rapidly evolve away from the desired parameters.  In \orb{} temperature gradient control and
injection of energy flux are imposed through sources added to the r.h.s. of the Vlasov
equation. These do not model the detailed physics of a realistic heat source (for example, the
temperature anisotropy generated by resonant heating schemes) but simply control moments of the
distribution function.

For the control of the temperature gradient, so that it stays close to an initial gradient, a {\it
  thermal relaxation operator} is used (this can be seen as an effective interaction with a heat
bath) of the form
\begin{equation}
  \label{eq:heatop2}
  S_{\text{H1}}  = -\gamma_\text{H} \left[ \delta f(\epsilon,s) - f_0(\epsilon,s) \frac{\overline{\delta f(\epsilon,s)}}{ \overline{f_0(\epsilon,s)}} \right],
\end{equation}
where the overbar is a flux-surface average. This source term maintains the distribution function
$f(\epsilon,s)$ close to the initial value, \ie{} it relaxes back to $f_0$ with a rate
$\gamma_\text{H}$. Note that the heating operator, \eqref{eq:heatop2}, does not act as a noise
control, unlike the modified Krook operator defined in Section~\ref{sec:Krook}. The second term in
the equation ensures that the gyrocenter density is not modified by the source term, \ie{} the heat
source does not act as an effective charge source. Due to the symmetry of this operator in $v_{||}$
it also does not add parallel momentum to the system; testing \cite{McMillan2008,McMillan2014} has
shown that long wavelength flows are largely unaffected by this heat source although certain higher
order effects could lead to significant flow drive on shorter wavelengths \cite{Sarazin2011}.

The choice of $\gamma_\text{H}$ determines how strongly the temperature gradient is clamped to the
initial gradient; since the form of the heating is not physical, it is necessary to set
$\gamma_\text{H}$ small enough not to excessively damp temperature corrugations; empirical
investigations suggest that $\gamma_\text{H}$ being ten times smaller than typical instability
growth rates is sufficiently small for convergence. It is possible to specify this heat source to be
active only in certain regions of the plasma, so that, for example, a ``source-free'' region in the
middle of the simulation domain may be obtained.

Fixed-input power simulations may be obtained by using a {\it fixed heat source} of the form
\begin{equation}
 S_{\text{H2}} =  \gamma_\text{R}(s) \frac{\partial  f_0}{\partial T},
\end{equation}
where $\gamma_\text{R}(s)$ is a spatial heating profile written in terms of an effective inverse
timescale over which the local temperature would vary in the absence of transport. Generally this
operator is used to represent a fixed input power source in the core of the tokamak. To model the
energy losses near the edge, two options can be chosen: first, to define a profile
$\gamma_\text{R}(s)$ with negative values in the edge region; second, to define a buffer region near
the boundary in which a Krook operator is specified (see Section \ref{sec:Krook}), thus damping the
edge profiles close to their initial values.

\subsection{Parallelization}
\label{sec:parallelization}
\begin{figure}[htbp!]
  \centering \includegraphics[width=0.7\textwidth]{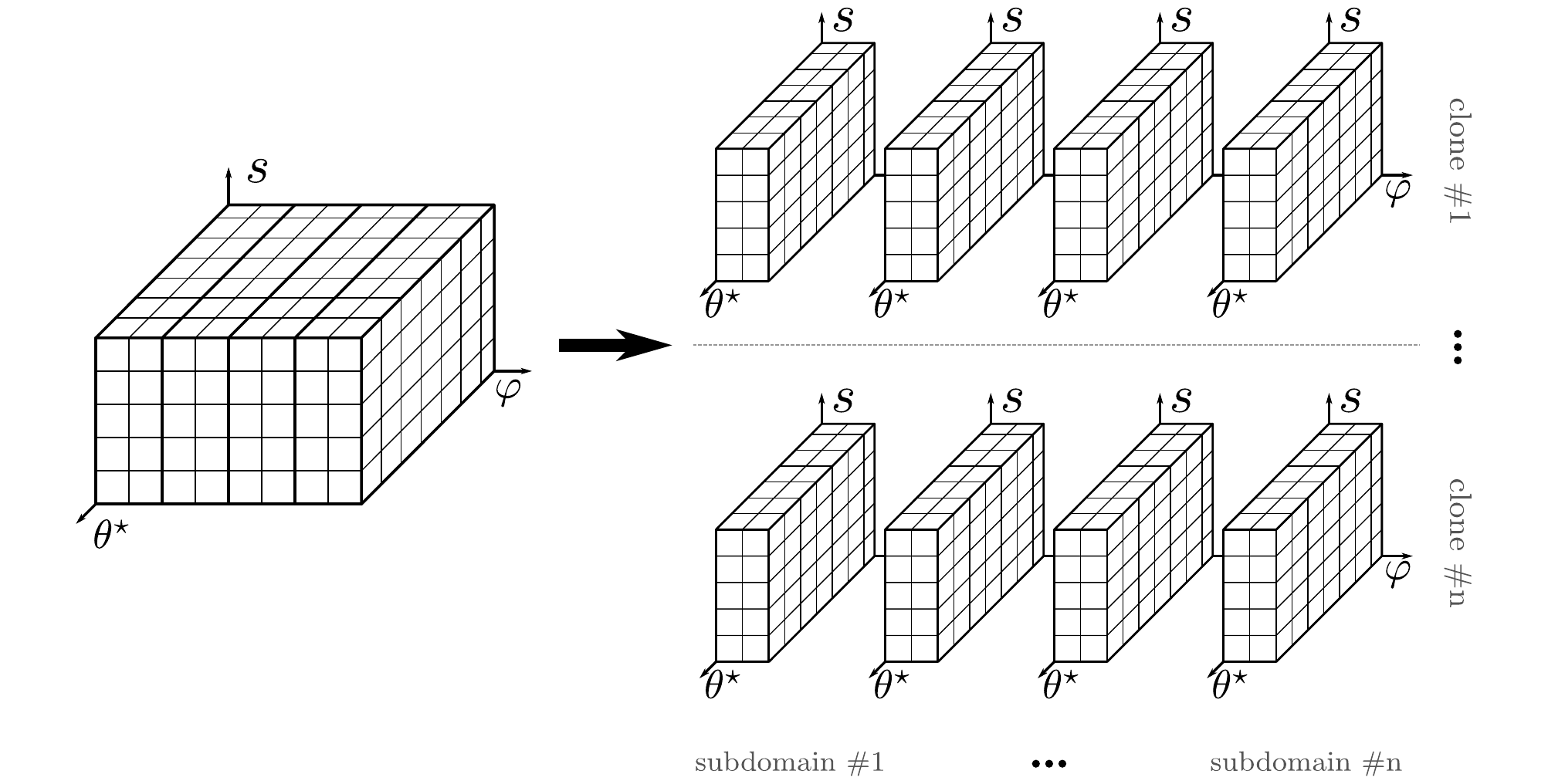}
  \caption{MPI parallelization using domain decomposition and domain cloning}
  \label{fig:domain_decomposition}
\end{figure}

In order to simulate complex physical systems in a reasonable amount of time, the \orb{} code is
massively parallelized using a hybrid MPI/OpenMP and MPI/OpenACC implementation. The MPI
parallelization is done using both domain cloning and domain decomposition \cite{Kim2000,Hatzky2006}
techniques, \figref{fig:domain_decomposition}.

The physical domain is first replicated into disjoint clones and the markers are evenly distributed
among them. Each clone can be further decomposed by splitting the physical domain in the toroidal
direction into subdomains. Each subdomain of each clone is attributed to an MPI task such that the
total number of processes is given by $P_{\text{MPI}} = P_{\text{sub}}\times P_{\text{clones}}$,
where $P_{\text{sub}}$ and $P_{\text{clones}}$ are respectively the number of tasks attributed to
the subdomains and clones.

After each time step, data must be transferred between the clones and subdomains. For the clones,
mainly global reductions of grid quantities are required, \eg{} after each charge deposition step
all the contributions from the clones must be gathered to compute the self-consistent
electromagnetic fields. For the subdomains, it consists of nearest neighbour communications for the
guard cells, global communications of grid data (parallel data transpose) for Fourier transforms and
point to point communications of particle data where we exchange the particles that have moved from
a subdomain to another. Note that in \orb{}, the particle exchange algorithm is not restricted to
the nearest neighbours, all-to-all is supported.

While the domain decomposition scales well with the number of subdomains, a large number of clones is
problematic in terms of performance. Indeed, the domain cloning approach is quickly limited by the
more demanding communications and the memory congestion due to the field data replication. To
overcome this issue each MPI task is multithreaded using OpenMP. This has the main advantage of
limiting the number of clones while still increasing the code performance by sharing the workload
among threads.

To take advantage of the new HPC platforms equipped with accelerators, the \orb{} code has been
recently ported to GPU using OpenACC. These developments will be detailed in a separate paper
\cite{Ohana2019}. The choice of using OpenMP and OpenACC was motivated because they allow us to keep
all options in a single source code version.

%%% Local Variables:
%%% mode: latex
%%% TeX-master: "../ORB5Review"
%%% End:

%% file: sections/results.tex
\section{Results}
\label{sec:results}

\subsection{Parallel scalability}
\begin{figure}[htbp!]
  \centering
  \includegraphics[width=18cm]{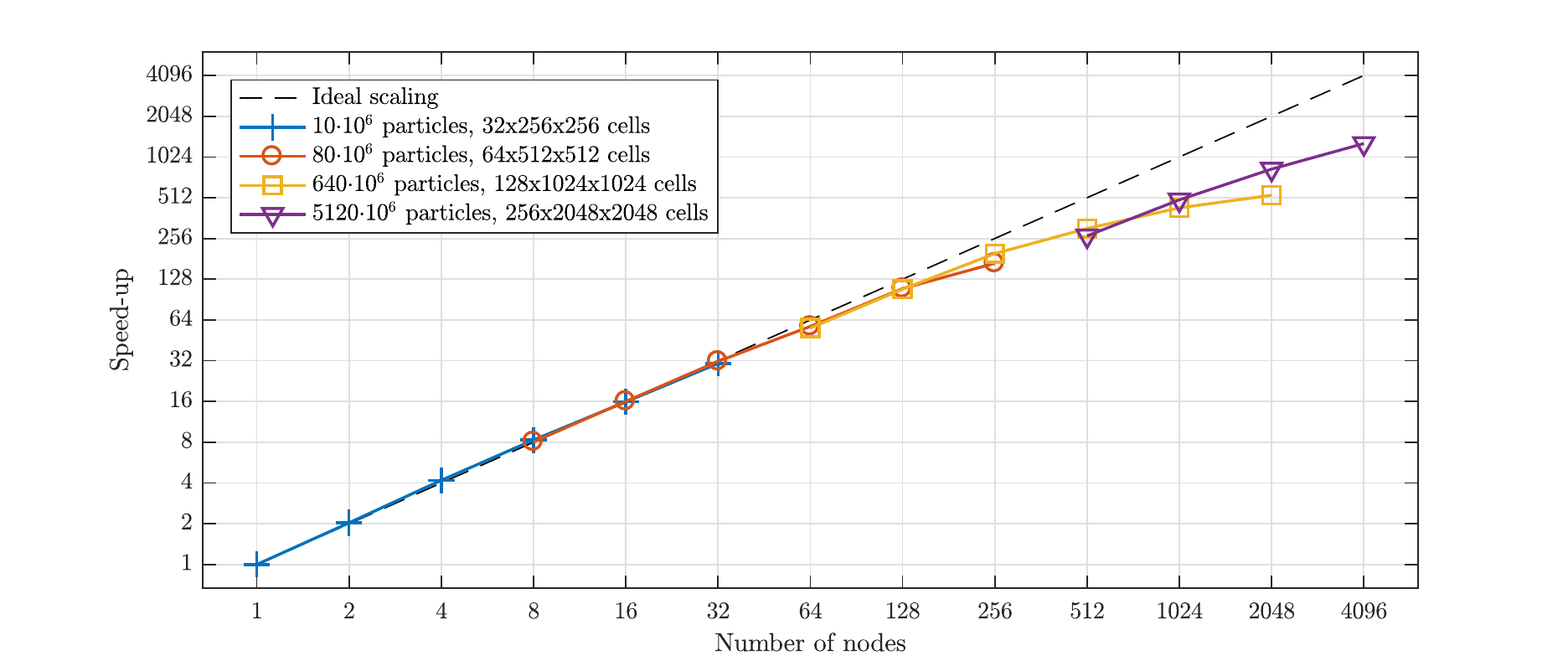}
  \caption{Strong scalings on the Piz Daint machine (Cray XC50, 12 cores per node). Speed-up is defined
    as inverse ratio of wall clock time to single node wall clock time, multiplied by powers of 8
    for the weak scaling factor. The number of particles indicates the number of ions and the number
    of electrons.}
  \label{fig:parallel_scalability}
\end{figure}

In \figref{fig:parallel_scalability}, we perform series of strong scalings of a typical
electromagnetic simulation with kinetic electrons. All the runs are made on the Piz Daint
supercomputer hosted at CSCS in Switzerland which is a hybrid Cray XC40/XC50 machine. For this
scaling, up to 4096 compute nodes of the XC50 partition equipped with one 12-core Intel Xeon E5-2690
v3 at 2.60GHz are used.

We use as many ions as electrons, using an adaptive number of Larmor points per guiding center going
from 4 to 32. The simulations are nonlinear, with a fixed number of 2 iterations for the control variate
scheme. Cubic splines are used. Scalar and 1D diagnostics are computed every other time step and 2D
diagnostics one step out of ten.

The starting point of each strong scaling makes a weak scaling where the grid resolution is
multiplied by 2 in each dimension, the number of particles by 8 and the number of compute nodes by
8. We use domain cloning inside nodes and domain decomposition in between them, meaning that the
number of clones is set to the number of cores per node, \ie{} 12, and the number of subdomains to
the number of nodes. We make an exception for the large-scale cases where the number of nodes
exceeds the number of toroidal cells, \ie{} the last points of the $640\cdot10^6$ particles and
$5120\cdot10^6$ particles cases, in which case we double the number of clones so that the number of
parallel tasks is equal to the product of subdomains and number of clones.

\orb{} scales very well up to 128 nodes with a speed-up larger than 85\% of the ideal speed-up. We
even get a small superscalability from 2 to 16 nodes thanks to increased data locality and decreased
memory congestion.

Above 256 nodes, the speed-up is limited mainly by the MPI communications of parallel data transpose
required for the field Fourier transforms. Some effort is currently put on reducing the cost of
those communications.

The performance of the GPU-accelerated \orb{} will be assessed in a following paper. In short, for
scaling tests similar to \figref{fig:parallel_scalability}, representative of production runs, the
GPU-accelerated \orb{} is up to 4 times faster than the CPU-only code.

\subsection{Strong flows and toroidal rotation}

We demonstrate the use of the strong flow features of the code using an adiabatic electron CYCLONE
benchmark case with nominal toroidal rotation rate $\Omega_R = 0.2 c_s/R$. The numerical parameters
are similar to those used for typical global CYCLONE benchmark cases with
sources\cite{Lapillonne2010a} (circular concentric equilibrium, $\rho*=1/180$, $a/R=0.36$,
plateau-like initial logarithmic temperature gradient profiles with $R/L_T=6.9$ and
$R/L_n=2.2$). The field solver grid is
$N_s \times N_{\ts} \times N_{\varphi} = 128 \times 512 \times 256$, and $1.2 \times 10^8$ markers
are used.  A heating operator is used with the rate $0.013 c_s/a$ to maintain temperature profiles near
their initial value. Coarse graining is applied every $2.8 a/c_s$ time units, with 64 bins in energy
and pitch angle, and a blending factor of $1$ (so all weights in a coarse-graining bin are set
equal).

\begin{figure}[!htbp]
  \centering
  \includegraphics[width=9cm]{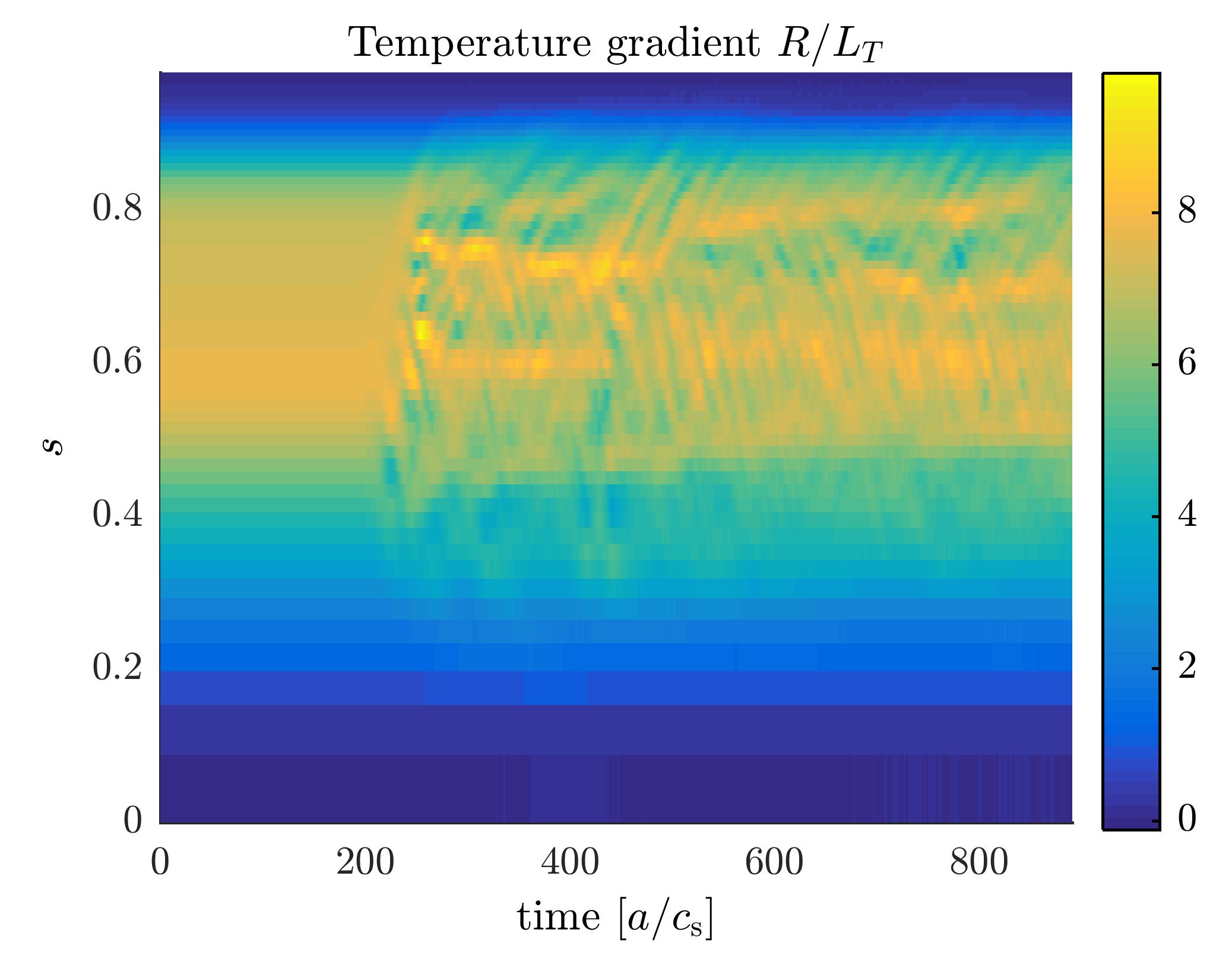}
  \caption{ Temperature gradient $ R / L_T $ versus time and radius in a strong flow simulation.}
  \label{fig:strongflow_rlt}
\end{figure}

\begin{figure}[!htbp]
  \centering
  \includegraphics[width=9cm]{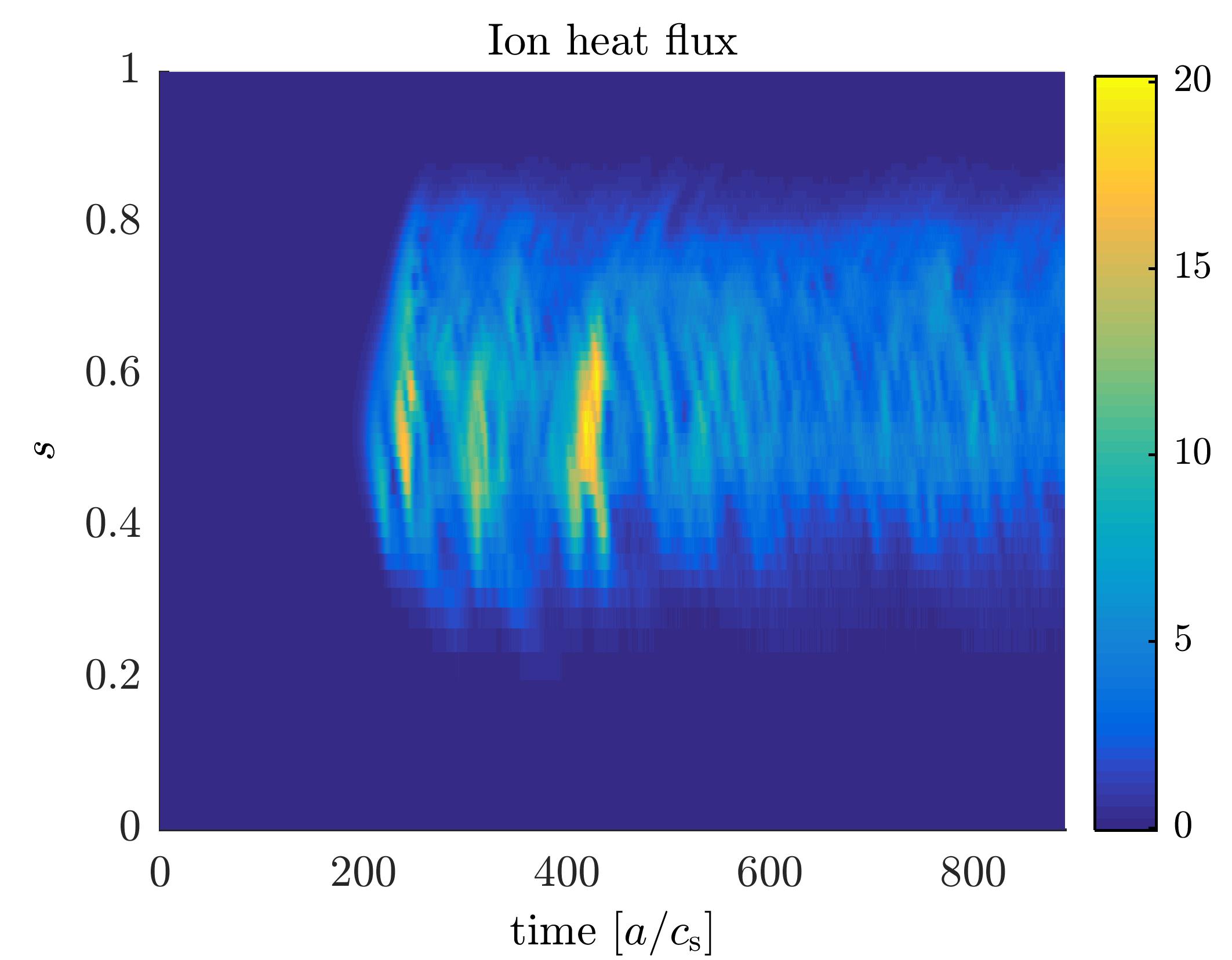}
  \caption{ Ion heat flux in gyro-Bohm units versus time and radius in a strong flow simulation.  }
  \label{fig:strongflow_flux}
\end{figure}

\begin{figure}[!htbp]
  \centering
  \includegraphics[width=9cm]{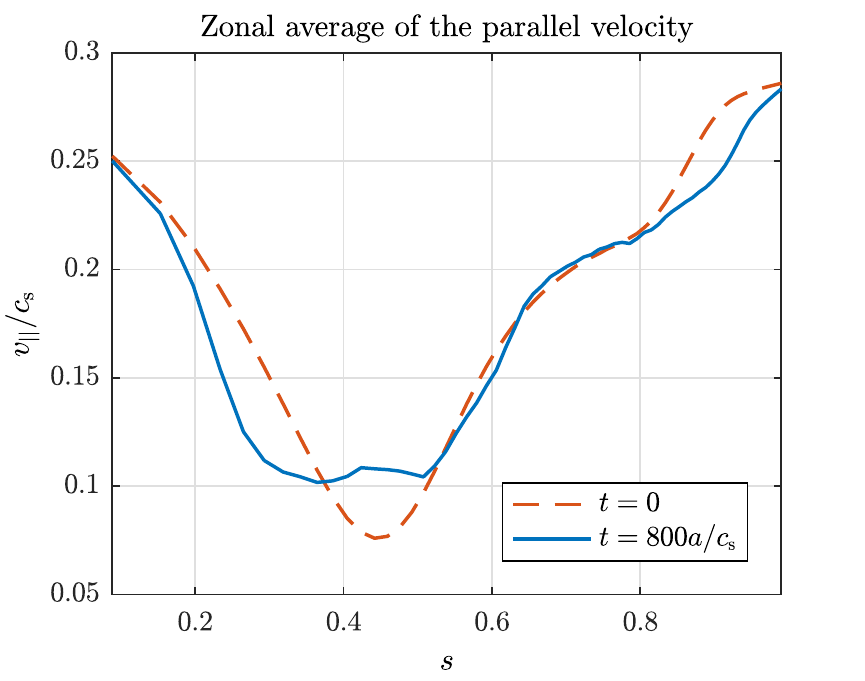}
  \caption{ Zonal average of parallel velocity versus radius at the beginning (red dashed trace) and
    end (blue trace) of a strong flow simulation.  }
  \label{fig:strongflow_pll}
\end{figure}

The effects of strong rotation on the equilibrium have been discussed earlier for
\orb{}~\cite{Collier2016}, so we focus on demonstrating the operation of the code in the nonlinear
regime; at the moderate levels of rotation tested here the effects are not expected to be
dramatic. As in non-rotating simulations, there is some overall relaxation of the heat profiles as
the turbulence driven transport commences,
Figs.~\ref{fig:strongflow_rlt}--\ref{fig:strongflow_flux}. The parallel flow profile,
Fig~\ref{fig:strongflow_pll}, is not constrained by the heating operator and relaxes slightly (note
that the initial parallel velocity profile is not completely flat, as might be expected for solid
body rotation). In these simulations, although strong flow effects due to Centrifugal and Coriolis
drift are included, the pinch driven momentum flux is expected to be nearly zero due to the use of
an adiabatic electron model: this is consistent with the observation of little net momentum flux in
these simulations.

\subsection{GK simulations of Alfv\'en modes in the presence of turbulence}
One of the main recent developments of \orb{} has been to allow for electromagnetic simulations. The
electromagnetic extension, via the control-variate scheme \cite{Mishchenko2014a,Hatzky2007} was
initially implemented in 2011 \cite{Bottino2011} and proved to work for ITG instabilities. Further
improvements have been necessary for having successful shear-Alfv\'en wave (SAW) tests, and the
first results have been published in 2016 \cite{Biancalani2016}. SAWs are known to be crucial in
present tokamaks and future reactors, as they can be driven unstable by energetic particles (EP),
which can deteriorate the EP confinement \cite{Chen2016}.

The linear dynamics of Alfv\'en instabilities investigated with \orb{} has been recently benchmarked
against analytical theory and other codes \cite{Biancalani2016,Konies2018a}. Moreover, the nonlinear
dynamics of Alfv\'en modes due to the wave-particle nonlinearity has been investigated with \orb{},
and compared with the GK code EUTERPE \cite{Cole2017,Biancalani2017}, where in particular a detailed
study of the saturation levels due to wave-particle nonlinearity has been carried out. Finally,
after a dedicated phase of verification and benchmarking on the Alfv\'en dynamics, \orb{} has now
started the investigation of the self-consistent interaction of Alfv\'en instabilities and
turbulence. Here, we describe a test case where the nonlinear dynamics of an Alfv\'en mode is
investigated in the presence of turbulence.

The tokamak geometry and magnetic field is taken consistently with Ref.~\cite{Biancalani2016}, for
the case referred to as energetic particle modes. Regarding the bulk profiles, the
ion and electron temperatures are taken equal everywhere, $T_\text{e}(s)=T_\text{i}(s)$. Here,
differently from Ref.~\cite{Biancalani2016}, a value of $T_\text{e}(s=\text{speak})$ corresponding
to $\rho^* = \rho_s/a = 0.00571$, is chosen. The electron thermal to magnetic pressure ratio
is $\beta_\text{e} = 5\cdot10^{-4}$.  An analytical function is used for the profiles of the equilibrium
density and temperature, for the three species of interest (thermal deuterium, labelled here as
``d'', thermal electrons, labelled here as ``e'', and hot deuterium, labelled here as ``EP''). For
the EP density, for example, the function is written as
$ n_{\text{EP}}(s)/n_{\text{EP}}(s_r) = \exp \curl-\Delta \, \kappa_n \tanh\bl(s-s_r)/\Delta\br\curr $.  The
value of $\Delta$ is the same for all species, for both density and temperature: $\Delta = 0.208$.
Deuterium and electrons have $\kappa_n=0.3$ and $\kappa_T=1.0$, and the EP have $\kappa_n=10.0$ and
$\kappa_T=0.0$. The EP temperature is given by $T_{\text{EP}}/T_\text{e} = 100$.  The distribution
function of the EP population is Maxwellian in $p_\|$. The EP averaged concentration is
$\left<n_{\text{EP}}\right>/n_e = 0.002$. A filter allows poloidal and toroidal mode numbers with $-128<m<128$
and $0\leq n<40$ to develop. Unicity boundary conditions are imposed at s=0.0 and Dirichlet at s=1.0. A
white noise initial perturbation is set at t=0. The electron mass is chosen as
$m_\text{e}/m_\text{i} = 0.005$. A Krook operator is applied to deuterium and electrons.

\begin{figure}[t!]
\begin{center}
\includegraphics[width=9cm]{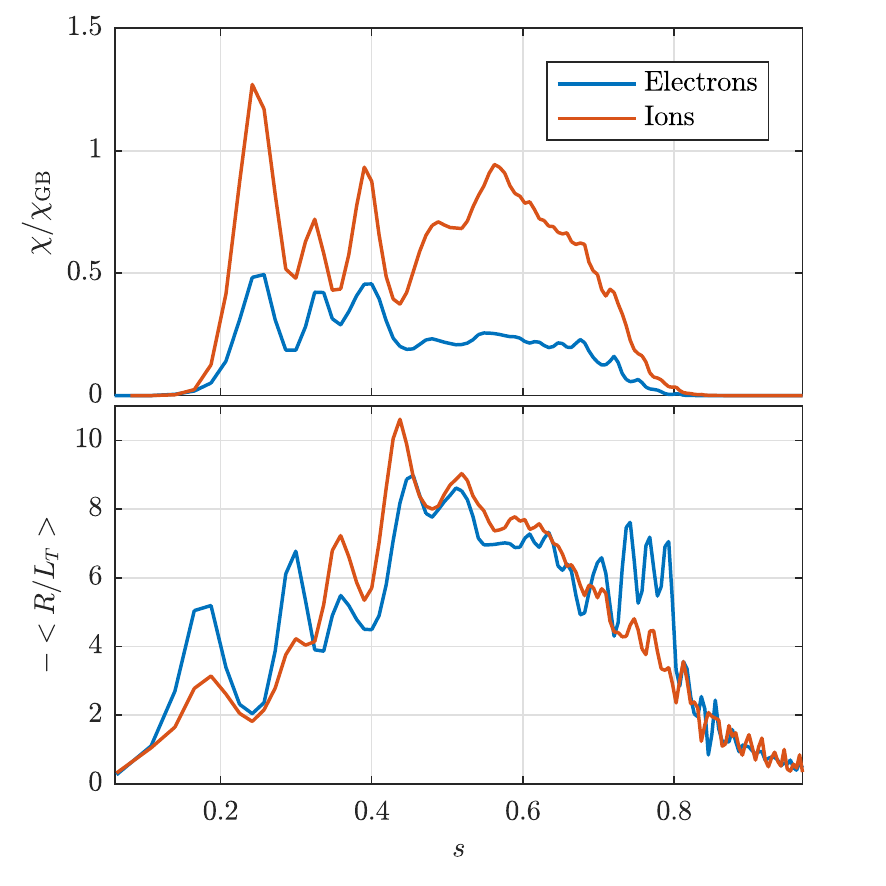}
\caption{\label{fig:chi_RLT_s} Corrugations of the effective heat diffusivity radial profile
  (top). The heat diffusivity is in gyro-Bohm units, $\chi_{\rm GB} =
  \rho_\text{s}^2c_\text{s}/a$. Radial profile of the temperature logarithmic gradient $R/L_T$
  (bottom). All the profiles are time averaged in $t\in[500, 1515]\ [a/c_\text{s}]$.}
\end{center}
\end{figure}

\begin{figure}[t!]
\begin{center}
\includegraphics[width=14cm]{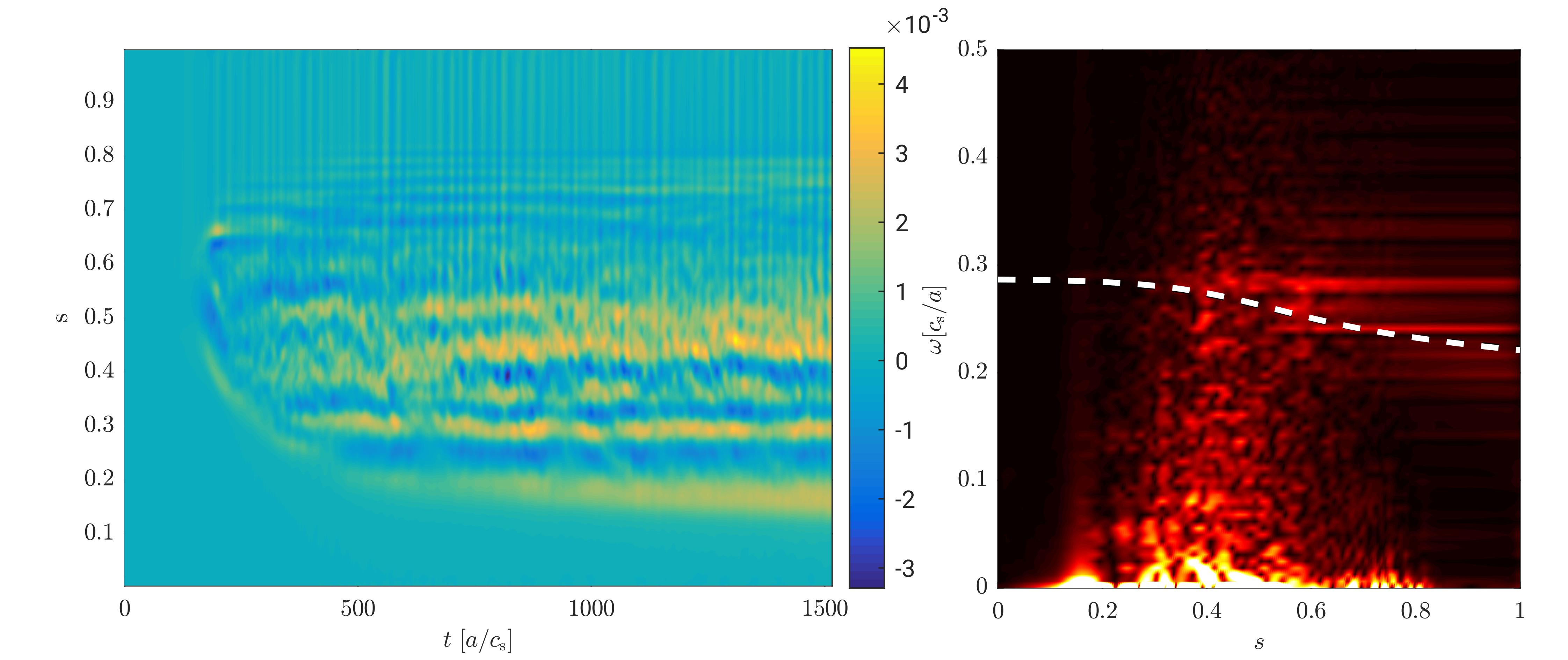}
\caption{\label{fig:dphibards}$E\times B$ velocity as a function of time and radius (lest) and the
  corresponding frequency spectrum (right). The white dashed line represents analytical estimates of
the GAM frequency \cite{Angelino2008}.}
\end{center}
\end{figure}

Nonlinear collisionless electromagnetic simulations have been performed with \orb{}, with turbulence
driven by the equilibrium temperature gradients, peaked at mid-radius with and without EP. In the
absence of EP, heat transport exhibits radial corrugated structures, \figref{fig:chi_RLT_s}; larger
corrugations are observed in the inner half of the radial domain, \ie{} $s\in[0, 0.5]$. Those
corrugations are also present as long-lived structures in the $E\times B$ velocity profile,
\figref{fig:dphibards} (left), and are particularly visible for $s\in[0, 0.5]$. Avalanches of
$E\times B$ velocity are generated at $s\sim0.5$ with a frequency matching the local GAM frequency,
\figref{fig:dphibards} (right), and then propagate outward with constant frequency. Finally, the
nonzonal component of the scalar potential has been measured, and observed to grow linearly in the
first so-called linear phase of the ITG turbulence, and then saturate.

\begin{figure}[t!]
\begin{center}
\includegraphics[width=0.48\textwidth]{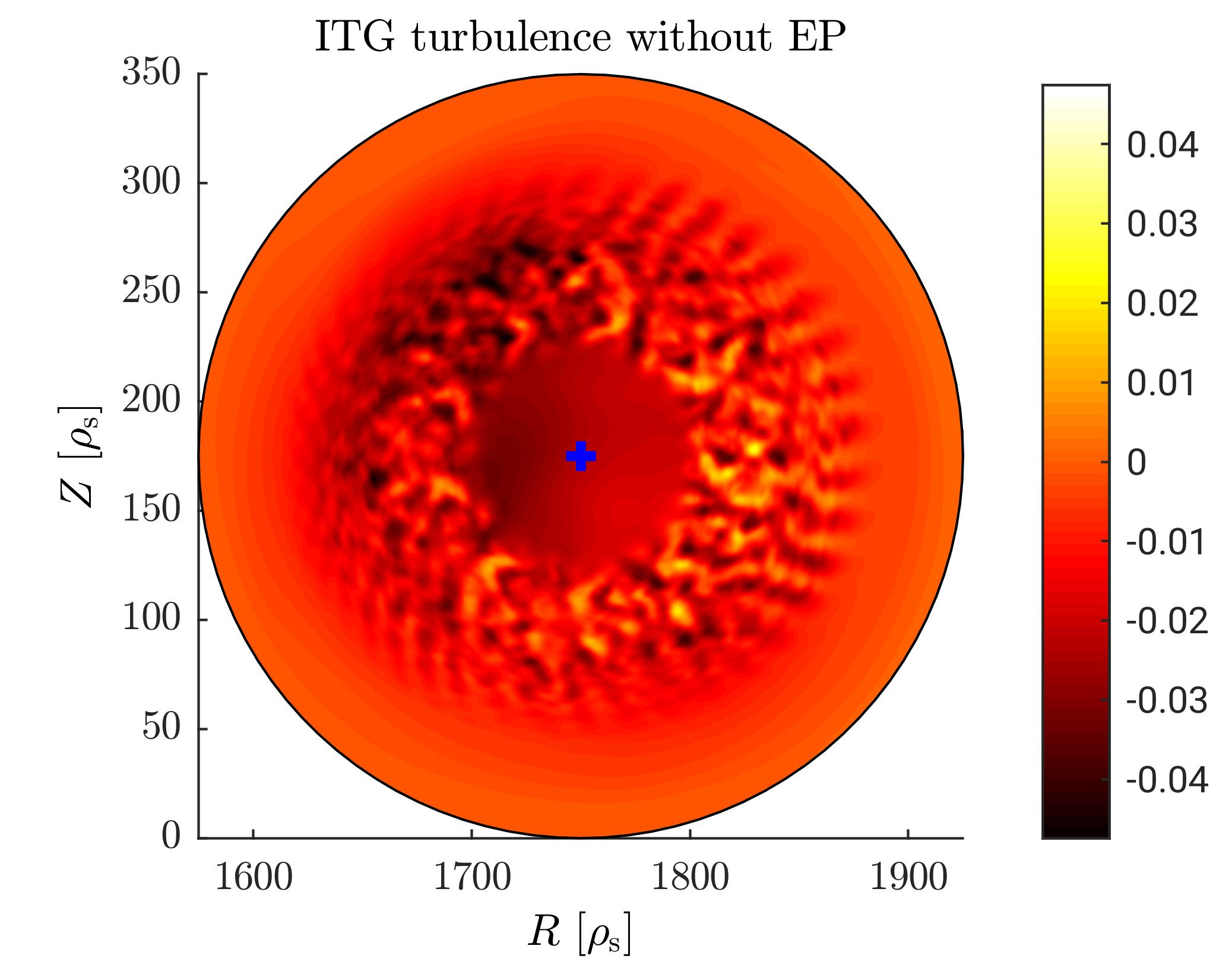}
\includegraphics[width=0.48\textwidth]{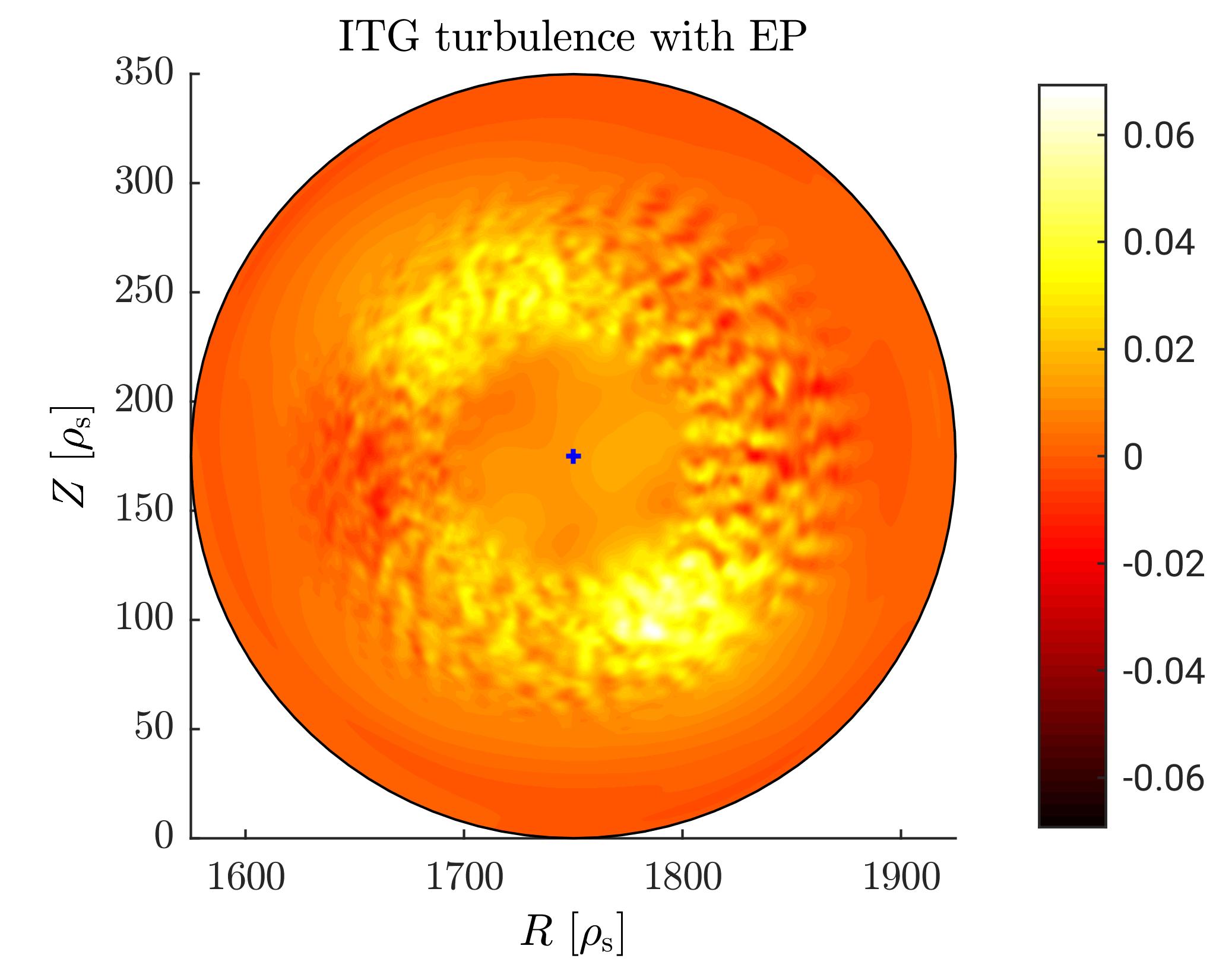}
\vskip -1em
\caption{\label{fig:SAWturbEP} Poloidal cut of the perturbed potential $\phi-\overline{\phi}$. On
  the left, a poloidal cut showing the ITG turbulence without EP. On the right, the characteristic
  poloidal structure of the Alfv\'en instability in the presence of turbulence is shown.}
\end{center}
\end{figure}

In the simulation where EPs are loaded, an Alfv\'enic instability is observed growing on top of the
turbulence. Zonal structures, like zero frequency zonal flows and geodesic acoustic modes, play the
role of mediators of small-scale turbulence and large-scale Alfv\'en modes. A comparison of the
perturbed electric potential with and without EP is shown in \figref{fig:SAWturbEP}.

Such simulations are numerically demanding due to the fact that they investigate intrinsically
multi-scale phenomena. Thus, high resolution in space and time is needed like in turbulence
simulations, the electrons must be treated kinetically for driving the current perturbations
necessary for the evolution of the Alfv\'en physics, and three separate plasma species (thermal
ions, thermal electrons, and EP) must be evolved simultaneously, in order to drive the Alfv\'en mode
unstable, with a corresponding high number of markers adopted.

%%% Local Variables:
%%% mode: latex
%%% TeX-master: "../ORB5Review"
%%% End:

%% file: sections/conclusion.tex
\section{Conclusion}
\label{sec:conclusion}
\orb{} is a global PIC code used to solve the electromagnetic gyrokinetic equations in presence of
collisions and various sources, \eg{} heat and strong flows. The Vlasov-Maxwell model on which it
relies is derived from variational principles and all the different physical approximations are
included in the gyrokinetic action. This allows to consistently derive the equations of motion while
ensuring conservation properties that can be later used to assess the simulation quality for
example.

Three models are available for the Poisson equation in which the ion polarization density can be
represented at full order, up to the second order in FLR corrections, or using a Padé
approximation. On the other hand, Ampère's equation is computed up to second order in FLR
corrections. While the ions are a gyrokinetic species, the electrons can be treated as adiabatic,
drift-kinetic, or an hybrid mix where passing electrons are adiabatic and trapped electrons are
drift-kinetic. Furthermore, the hybrid model, which does not respect the ambipolarity condition, has
been corrected.

The code is based on the PIC $\delta f$ control variate scheme in order to reduce the numerical noise
due to finite particle sampling and various other techniques are used to further limit this noise,
\eg{} noise reduction schemes are implemented to constrain the particle weight spreading and a
Fourier filter allows to solve only the physically relevant modes. The \orb{} code is parallelized
using an hybrid OpenMP/MPI or OpenACC/MPI approach allowing to benefit from the many and multicore
HPC systems. Scalability experiments have shown \orb{}'s excellent parallel scalability up to
thousands of cores.

The \orb{} code has been carefully and extensively benchmarked against various Lagrangian and
Eulerian gyrokinetic codes and always showed a good agreement in the results
\cite{Lapillonne2010a,McMillan2010a,Gorler2016,Biancalani2017a,Konies2018a,Merlo2018a}. A few
physical simulations run with the \orb{} code and including \eg{} strong flows, toroidal rotation,
and shear Alfvén waves are also presented to illustrate the capabilities of the code.

\section{Acknowledgments}
The authors deeply acknowledge past valuable contributions of Dr. T. M. Tran to the \orb{} code
development. Fruitful discussions with S. Ethier are gratefully acknowledged.

The authors thank the CSCS for providing access to the full Piz Daint machine to perform the
scalability tests. This work was supported by a grant from the Swiss National Supercomputing Centre
(CSCS) under project ID s760. This work has been carried out within the framework of the EUROfusion
Consortium and has received funding from the Euratom research and training programme 2014-2018 and
2019-2020 under grant agreement No~633053. The views and opinions expressed herein do not
necessarily reflect those of the European Commission. This work was partly supported by the Swiss
National Science Foundation.

%%% Local Variables:
%%% mode: latex
%%% TeX-master: "../ORB5Review"
%%% End: